\documentclass[lettersize,journal]{IEEEtran}
\usepackage{amsmath,amsfonts}
\usepackage{amsmath,amssymb,amsfonts}
\usepackage[ruled]{algorithm2e}
\usepackage{array}
\usepackage[caption=false,font=normalsize,labelfont=sf,textfont=sf]{subfig}
\usepackage{textcomp}
\usepackage{stfloats}
\usepackage{url}
\usepackage{verbatim}
\usepackage{graphicx}
\usepackage{cite}
\usepackage[flushleft]{threeparttable}
\usepackage[utf8]{inputenc}
\usepackage{booktabs}
\usepackage{float}
\usepackage{csquotes}
\usepackage{times}
\usepackage{url}
\usepackage{array,color}
\usepackage{multirow}
\usepackage{soul}
\usepackage[table]{xcolor}
\usepackage{lastpage,fancyhdr,graphicx}
\usepackage{epstopdf}
\usepackage{rotating}
\usepackage{bm}
\usepackage[colorinlistoftodos]{todonotes}

\DeclareMathOperator*{\argmax}{\arg\!\max}

\SetCommentSty{mycommfont}
\newcolumntype{P}[1]{>{\centering\arraybackslash}p{#1}}
\usepackage{xr}
\usepackage[colorlinks]{hyperref}
\newtheorem{definition}{Definition}[section]

\usepackage{enumitem}
\newcounter{descriptcount}

\begin{document}
\title{Fusing Interpretable Knowledge of Neural Network Learning Agents For Swarm-Guidance}
\author{Duy~Tung~Nguyen,~\IEEEmembership{Student Member,~IEEE,}
        Kathryn~Kasmarik,~\IEEEmembership{Senior Member,~IEEE,}
        and~Hussein~Abbass,~\IEEEmembership{Fellow,~IEEE}
\thanks{The authors are with the School of Engineering and Information Technology, University of New South Wales, Canberra, ACT, 2600.\protect\\
E-mail: tung.d.nguyen@unsw.edu.au}
\thanks{Manuscript received xxxxxx; revised xxxxxx.}}
\markboth{Journal of \LaTeX\ Class Files,~Vol.~14, No.~8, August~2021}%
{Shell \MakeLowercase{\textit{et al.}}: A Sample Article Using IEEEtran.cls for IEEE Journals}


\maketitle

\begin{abstract}
Neural-based learning agents make decisions using internal artificial neural networks. In certain situations, it becomes pertinent that this knowledge is re-interpreted in a friendly form to both the human and the machine. These situations include: when agents are required to communicate the knowledge they learn to each other in a transparent way in the presence of an external human observer, in human-machine teaming settings where humans and machines need to collaborate on a task, or where there is a requirement to verify the knowledge exchanged between the agents. We propose an interpretable knowledge fusion framework suited for neural-based learning agents, and propose a Priority on Weak State Areas (PoWSA) retraining technique. We first test the proposed framework on a synthetic binary classification task before evaluating it on a shepherding-based multi-agent swarm guidance task. Results demonstrate that the proposed framework increases the success rate on the swarm-guidance environment by 11\% and better stability in return for a modest increase in computational cost of 14.5\% to achieve interpretability. Moreover, the framework presents the knowledge learnt by an agent in a human-friendly representation, leading to a better descriptive visual representation of an agent\textquoteright s knowledge.
\end{abstract}


\section{Introduction}\label{jpaper2-introduction}

\IEEEPARstart{C}{onventional} transfer learning research~\cite{pan2009survey,weiss2016survey} focuses on the adaptation of a model to a target task that may have faced a drift in its underlying distribution. The drift may occur in the sources of data, the mapping from the data to the decision/class, or class distribution. An inter-task knowledge transfer process often takes into account similarity or relevance among tasks concerning several factors, including the objectives and complexity of the task, to aid a model\textquoteright s training~\cite{zhang2018overview}.

In practice, the transfer of knowledge requires an agent architecture and infrastructure, where the knowledge to be fused needs to be packaged into appropriate messages to be transmitted from the sender to a receiver agent. This process imposes additional constraints ranging from classic constraints imposed by bandwidth limitation or the maximum length of a message due to the communication protocol, to contemporary constraints imposed by the system\textquoteright s owner including the need for these messages to be in a language suitable for the system to check their integrity and for an external human to comprehend. When an agent\textquoteright s internal representation of knowledge takes a complex mathematical form such as a neural network, with a large number of nonlinear transformations occurring in parallel to approximate the underlying function required to perform a particular task, it becomes pertinent that an agent may need to use a different representational language, such as ``IF $\dots$ THEN $\dots$" rules, to transfer knowledge from the internal representation of knowledge, such as a neural network. This latter scenario is the context of this paper, whereby we present a framework that allows agents to use neural networks for action production and a conditional syllogism for knowledge transfer.

In our previous work~\cite{nguyen2020towards}, we designed exact transformations from feedforward neural networks with ReLU activation functions to univariate and multivariate decision trees and conditional rules. In this paper, we use these transformations as the underlying internal mappings performed by an agent to transform their neural networks into conditional syllogisms. We then propose an interpretable knowledge fusion framework. 

The knowledge fusion process operates on decision rules as knowledge representation extracted from neural networks with a knowledge interpretation algorithm using the mathematical transformations presented in~\cite{nguyen2020towards}. Subsequently, the knowledge assessment and integration steps identify the valuable knowledge subsets and combine them with the original knowledge of the receiver respectively. We present an inverse algorithm that transforms the received rule-based representation back to a neural representation. We present a retraining procedure that allows the knowledge fusion module to integrate the received knowledge to accelerate the agent\textquoteright s adaptation process in previously unobserved environments. This retraining procedure is shown to be more efficient than adapting to the new environment through grass-root learning. Compared to the baseline knowledge transfer framework between multiple tasks, our proposed framework is not only maintained an equivalent or even higher performance but also provides a more interpretable form of representation for an external human observer. This paper is devoted to the fusion module to demonstrate the retraining algorithm and answer a few fundamental questions in the retraining process.

In Section~\ref{jpaper2-relatedwork}, we review relevant work on transfer learning for neural networks and inter-agent knowledge transfer frameworks. We then present the concepts and formal definition of the knowledge fusion problems in Section~\ref{jpaper2-problemdefinition}, followed by the proposed framework and the approaches for knowledge fusion in Section~\ref{jpaper2-methodology}. The experiments and results are presented in Sections~\ref{jpaper2-experiments}. Conclusions are drawn in Section~\ref{jpaper2-conclusion}, including a discussion of the limitations of the framework.

\section{Background Materials}\label{jpaper2-relatedwork}

There are a vast number of studies on transfer learning in the literature. In the domain of transfer learning, there are three major research areas: domain adaptation, multi-task transfer learning, and agent-agent knowledge transfer. Prior studies in the third area of research will be reviewed, as this is the emphasis of our study here.

Network-based transfer learning affords an agent the ability to adapt to new environments by bootstrapping from previously learnt knowledge. Research in this area can be categorised into three main categories. The first assumed a single large network held by each agent and focuses on how to map knowledge between these networks. The second uses modular neural networks by specialising a neural network for a particular sub-task; thus, decomposing the learning and knowledge transfer problems into smaller chunks. The third category relies on an interactive transfer framework between agents using Teacher-Student Learning based on an assumption that both agents do not know the other\textquoteright s type of knowledge representation. 

In the first category, Oquab et al.~\cite{oquab2014learning} transfer a part of a trained convolutional neural network (CNN), which is used originally to classify data in ImageNet, to a new network structure for estimating the visual characteristics of a different image dataset. The newly trained CNNs can learn object detection problems where the number of training samples is relatively low. Long et al.~\cite{long2016unsupervised} proposed a technique to simultaneously train adaptive models for unsupervised feature learning and approximate the residual function for the mapping between the source and target domains. Other studies also presented network-based transfer methods for adapting useful relevant features for different applications, ranging from detecting vehicles in video streams to detecting abnormalities in medical images~\cite{wang2018vehicle,zhu2016deep,hassan2020developing}.

In the modular neural networks category, a range of methods combines the power of many sub-networks, each of which is responsible for a sub-task in the problem. The approach is useful in complex tasks where a single non-modular network does not necessarily learn well. The fusion of all modules in an aggregate network better manages task complexity and improves generalization~\cite{hu2020classifier}. Modular neural networks have been applied to many problems including system modelling, pattern recognition, and prediction~\cite{hu2020classifier,qiao2018design,li2019design}. The use of modular neural networks facilitates inter-agent transfer as its sub-networks are trained to adapt to specific requirements or situations~\cite{tommasino2016reinforcement}. In~\cite{rajendran2015attend}, an attentive architecture called Attend, Adapt, and Transfer (A2T)  is introduced to leverage the knowledge from previous tasks to train agents on a new task. A2T combines the outputs of a new network and multiple pre-trained neural networks using an array of weights, which are generated by an attention network. Recently, a state-of-the-art model, called MULTIPOLAR~\cite{barekatain2020multipolar}, is designed based on a similar attentive approach. The difference lies in the use of a trainable aggregation weight matrix that is applied on source networks\textquoteright \ outputs element-wise instead of a simple linear combination between outputs. The use of a weight matrix is more lightweight than a whole attentive network and also more stable than the state-dependent attention mechanism in A2T. In general, the limitation of all modular neural networks is the size of the trained network is often large as multiple modules are included.

In the interactive learning category, Silva et al.~\cite{da2020agents} identify the key components in an inter-agent teaching scenario, called Teacher-Student Learning. They propose a framework in which multiple agents maintain a communication protocol via which agents can learn useful skills with guidance. Two types of knowledge transfer frameworks are introduced, including Teacher-Driven and Learner-Driven, which differentiate from each other by deciding who will play an active role in initiating the communication. The advice process can be used to modulate the learner\textquoteright s actions directly or through the learned values~\cite{cruz2016training}. These frameworks are applicable for knowledge transfer in a multi-agent setting or multi-task learning. However, the retraining processes in those studies are still expensive in the cases of skill exchange between agents as they have not fully taken advantage of the values of the learned experiences about the inter-task relevance. Moreover, they allow an implicit transfer of knowledge and do not provide a form of interpretable or verifiable knowledge representation.

Some of the challenges that knowledge transfer needs to overcome include the requirement of a large number of training data instances and computational resources to optimise the learning model and decision-making processes of an agent. A reinforcement learning agent, for example, requires a significantly large amount of data to learn a behaviour~\cite{yu2018towards}. This could be prohibitive for real-world problems due to the costs and risks associated with these explorations. Taking advantage of the experiences acquired from other already trained agents in similar scenarios is, therefore, a worthwhile endeavour~\cite{da2020agents}. Nevertheless, blind transfer of experiences could become counter-productive, as it neither helps designers to validate the knowledge being represented nor the knowledge being exchanged. When knowledge is learned with a neural network representation, the exchange of neural substrates could be a random guess that inhibits, rather promote, better learning. Therefore, an inter-agent knowledge transfer framework where the agents can proactively select and exchange relevant and significant knowledge is needed. Also, one other important requirement of a real-world system is that the exchange of knowledge between agents needs to be verifiable by another party so that they can evaluate whether the process is reliable~\cite{he2021challenges}. Such transparent knowledge transfer begs for mathematical transformations of the knowledge being transferred to interpret black-box models, such as neural networks, into a form that could be verified; allowing each agent to transmit relevant information alone.

\section{Problem Formulation}\label{jpaper2-problemdefinition}

In an agent-agent interaction setting, each individual can seek to interface with the other for communicating the objectives of a task, establishing collaboration, asking for advice, or sharing knowledge/skills to solve a task. In this paper, we focus on knowledge sharing between agents.

When agents attempt to share knowledge among themselves, it is pertinent to choose an appropriate common representation to communicate information and to establish a common medium for the interface between agents. Regularly, given different mechanism including the use of structures, processes or machine learning algorithms, each agent might store and processes its knowledge in a unique way. This unique way involves the encoding of the knowledge in a representational space of either explicit or implicit concepts and axioms, which reduces the accessibility and interpretability for agents with different cognition. Unfortunately, even two agents using the same type of models or representations of knowledge do not necessarily share the same spaces as they work on a problem. For example, two neural networks trained on the same problem (even the same data set but with different initializations), used by two agents respectively, could end up possessing very different weights. This leads to the different spaces that the input data will be mapped to. Therefore, a knowledge-sharing process might involve two sub-processes called knowledge interpretation and knowledge fusion. 

Knowledge interpretation is a process that maps the knowledge of an agent from one representation language to another~\cite{abbass2021symbiomemesis}. Consider a system of two agents performing knowledge sharing: sender agent $\pi$ and receiver agent $\beta$ with knowledge representations $\mathbb{K}_\pi$ and $\mathbb{K}_\beta$ respectively. Knowledge interpretation involves the transformations: $\mathbb{I}_{\beta\to\psi}: \mathbb{K}_\beta \to (\mathbb{K}_\beta)_\psi$ and $\mathbb{I}_{\pi\to\psi}: \mathbb{K}_\pi \to (\mathbb{K}_\pi)_\psi$ to unify the representation for later fusion.

Definition~\ref{def:knowledgefusion} formulate the problem of knowledge fusion, which is the focus of this paper. Assume the agent $\beta$ needs to solve a task $\mathcal{M}_j \in \mathbb{M}$, where $\mathbb{M}$ is the set of available tasks. An agent $\pi$ possessing certain knowledge on the task $\mathcal{M}_i \in \mathbb{M}$ is selected to send its knowledge to agent $\beta$ to help it well-adapt to its assigned task. Note that $\mathcal{M}_i$ and $\mathcal{M}_j$ can be the same task. The knowledge fusion problem can then be defined as below.

\begin{definition}[Knowledge Fusion Problem]\label{def:knowledgefusion} \emph{The knowledge fusion problem is to identify the best functional mapping to combine received knowledge with an agent\textquoteright s internal knowledge in such a way that maximizes the performance of the receiving agent $\mathbb{K}^{\mathcal{M}_j}_\beta$ on the task $\mathcal{M}_j$, that is,}
\begin{equation}
    {\mathbb{K}^{\mathcal{M}_j}_\beta}^* = \argmax_{\mathbb{K}^{\mathcal{M}_j}_\beta} \Upsilon(\mathbb{K}^{\mathcal{M}_j}_\beta) \;\; | \;\;\mathbb{K}^{\mathcal{M}_j}_\beta = \Theta((\mathbb{K}_\beta)_\psi,(\mathbb{K}_\pi)_\psi)
\end{equation}
where $\Upsilon(\mathbb{K}^{\mathcal{M}_j}_\beta)$ is the objective function of agent $\beta$ on task $\mathcal{M}_j$ after knowledge fusion. The objective function can be the performance of the agent or the probability of the agent with new knowledge achieving the highest performance after some further refinement of the knowledge. 
\end{definition}

It is worthwhile to state that the performance of an agent could combine multiple performance indicators including effectiveness, efficiency, and computational cost.

\section{Methodology}\label{jpaper2-methodology}

This paper builds on the knowledge interpretation algorithm presented in~\cite{nguyen2020towards}. We summarise that algorithm in the supplementary materials (Section~\ref{jpaper2-methodology-ECDT}) for completeness.  The knowledge interpretation algorithm converts the neural network representation into decision rules. Hence, there are two forms of knowledge representations: neural and rule forms. Neural networks are the original form of knowledge representations used by the agents for action production, while the rule form is more interpretable and suitable for communication and knowledge fusion, allowing agents to validate the new knowledge and for an external observer, if available, to comprehend and check it. 

\begin{figure}[!b]
    \centering
    \includegraphics[width=0.95\linewidth]{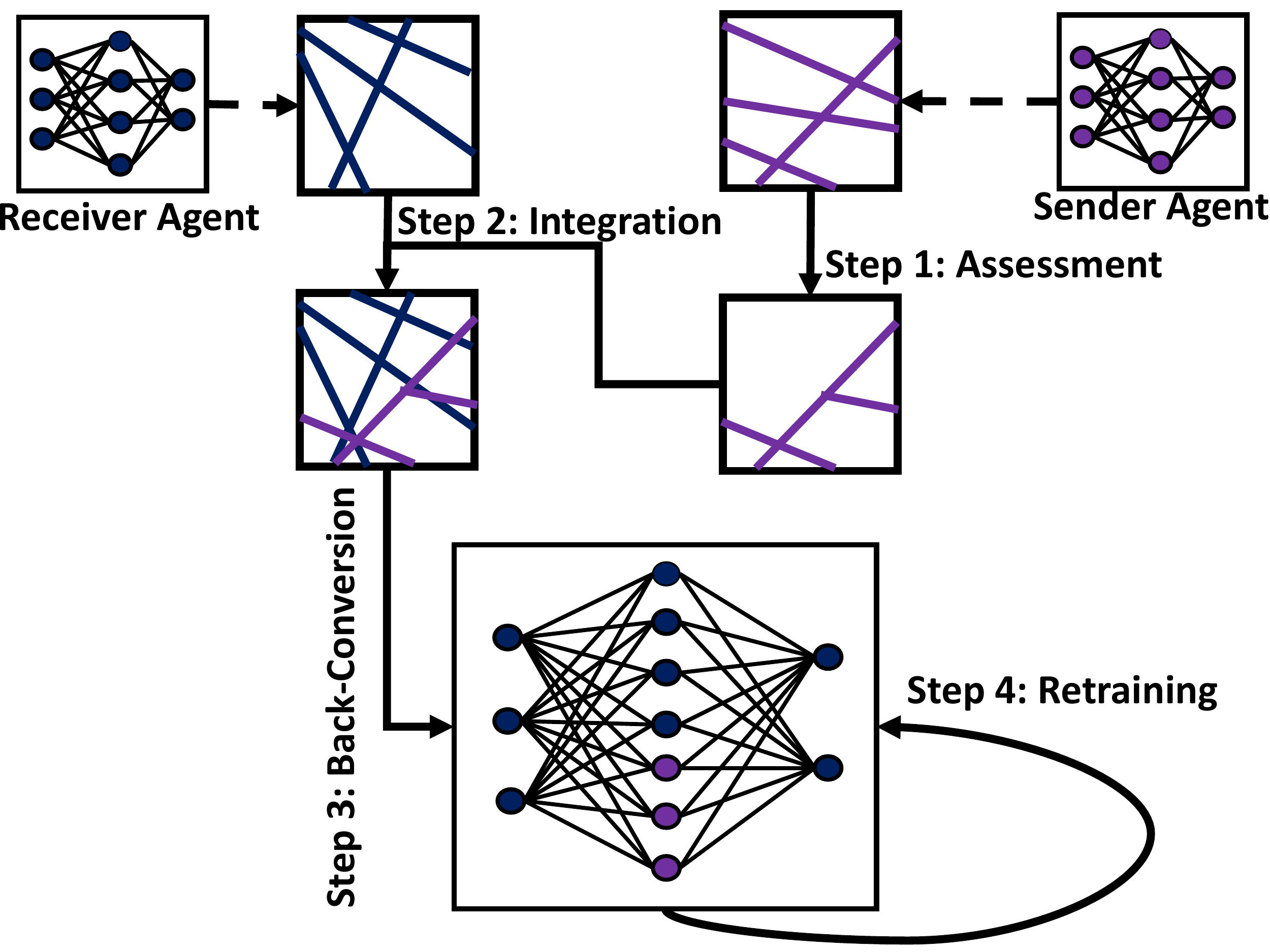}
    \caption{The proposed knowledge fusion framework.}
    \label{fig:knowledgefusion}
\end{figure}

In this paper, we focus on the knowledge fusion problem between two neural network agents. Our proposed interpretable knowledge fusion framework between neural network-based learning agents is illustrated in Figure~\ref{fig:knowledgefusion}.

Following are the steps of the proposed knowledge fusion process:
\begin{enumerate}
    \item \textbf{Assessment}: Evaluate the outputs of original and other agent\textquoteright s rules on the new target problem. Keep rules with higher performance than the original ones. Identify the remaining weak pieces of knowledge - the knowledge represented by the regions of state space that are directly linked with the failures.
    \item \textbf{Integration}: For rule forms, each rule geometrically covers a region in the state space. Find all overlaps between rules of the agent $\pi$ and agent $\beta$ and create a new set of rules from the overlaps.
    \item \textbf{Back-conversion}: Convert selected integrated rules from agent $\pi$ into neural substrates. Then, fuse the neural network of the agent $\beta$ with neural substrates from the agent $\pi$.
    \item \textbf{Re-training (Optional)}: Select priority levels for sampling data in weak/strong regions of state space represented by interpretable form, and retrain the network.
\end{enumerate}

The framework provides a way to fuse the knowledge of the sender and the receiver agents to appropriately bootstrap the latter agent in new, but similar to previously seen, situations/tasks. The receiver agent may need to continue training its neural network after receiving new knowledge to improve its skills in the new environment, especially if the new environment differs from the previous environments it was exposed to. We spend the remainder of this section describing each step in more detail.

\subsection{Rule Assessment}\label{jpaper2-methodology-fusion-assessment}
The rules, which can be extracted from the neural networks using the EC-DT algorithm presented in~\cite{nguyen2020towards}, are sets of multivariate inequalities. Each rule/set of inequalities extracted from the EC-DT tree geometrically represents the hyperplanes that constitute a polytope covering an area in the state space. This set of inequalities is called the H-representation of the polytope. Another form of representation of such polytope is called V-representation, which is a set of vertices of the polytope.

In this phase of the knowledge fusion step, a set of data are sampled from the target problem that the agent needs to adapt to. The data samples are fed to the sender and receiver\textquoteright s networks, and their performance is recorded. Each subset of sampled data belongs to a specific polytope in the ruleset of the sender or the receiver. If the performance of the model on data in a sender\textquoteright s polytope is higher than that of the receiver\textquoteright s polytope, then the sender\textquoteright s polytope is valuable. The piece of knowledge represented by that polytope should be merged into the knowledge base of the receiver agent. Otherwise, if the knowledge represented by a sender\textquoteright s polytope does not contribute to a better performance in the new problem space, it should not be merged into the receiver\textquoteright s knowledge base as it will unnecessarily increase the size of the memory used. The polytopes that cover the regions containing the samples that the original models underperform are identified as the weak areas of the knowledge. These areas should be prioritized in the re-training process at a later stage.

\subsection{Knowledge Integration}\label{jpaper2-methodology-fusion-transfer}
Before implementing the necessary integration of knowledge between agents, we first need to determine the overlapping regions between polytopes extracted from the sender, which are kept after the evaluation phase, and the polytopes from receiver agents in \textbf{Step 1} of the framework.  Algorithm~\ref{Alg:check_intersection} is used to check whether there is any intersection/overlap between two given polytopes. In this algorithm, the sets of inequalities, representing polytopes $P_{R_k}$ and $P_{S_j}$ corresponding to the knowledge chunks from receiver and sender agents, are converted to V-representations using the Double Description Method~\cite{motzkin20163}. The Python package for the implementation of the transformation between the H-representation and the V-representation is available from the website: \url{https://pypi.org/project/pycddlib/}.
Two polytopes are overlapping with each other when at least one of the vertices of a polytope is an interior point of the other polytope and vice versa.

\begin{algorithm}[!t]
\footnotesize
\SetKwInOut{Input}{Input}
\SetKwInOut{Output}{Output}
\Input{Polytopes $P_{R_k}$ and $P_{S_j}$}
\Output{The state of whether the overlap exists between two polytopes: $\{True, False\}$}

Convert convex polytope $P_{R_k}$ into list of corresponding vertices $V_{R_k}=\{v_{R_k}^1,v_{R_k}^2,...\}$  \\
\uIf{$\exists v_{R_k}^i \in V_{R_k} \; \mbox{such that} \; v_{R_k}^i \; \mbox{is an interior point of} \; P_{R_k}$}
{
	\textbf{return} $True$
}
\Else
{
	\textbf{return} $False$
}

\caption{Check intersection between two polytopes.}\label{Alg:check_intersection}
\end{algorithm}

When two polytopes are determined to have an intersection with each other, we have to find the set of inequalities that characterizes the overlapping region of those two. A simple way to determine this set of inequalities is to merge both sets of inequalities. However, this might result in a redundant set of inequalities, which is not optimized in terms of performance. In our framework, we employ a hyperplane redundancy removal technique with Clarkson\textquoteright s algorithm~\cite{clarkson1994more} (Algorithm~\ref{Alg:redundancy_removal}). In this algorithm, each set of hyperplanes goes through several loops, each loop determines whether a hyperplane/inequality is redundant to the final polytope. To check whether an inequality is redundant or not, we have to solve the following linear programming (LP) problem:

\begin{quote}
\textit{For a polytope formed by the set of inequalities $\{A_jX \leq b_j\ |\forall j \in J\}$, where $J$ is the set of indices for the corresponding inequalities, the inequality $A_kX \leq b_k$ is essential if the Linear Programming problem below has an optimal solution whose value is greater than $b_k$:}
\end{quote}

\vspace{-0.5cm}
\begin{align}
Test(J,k): \nonumber \\
& \mbox{maximize} \;\; A_k X \nonumber \\
& \mbox{subject to} \nonumber \\
& \qquad \qquad A_j X \leq b_j, \forall j \in J \setminus \{k\} \nonumber\\
& \qquad \qquad A_k X \leq b_k + 1 \nonumber
\end{align}

In layman\textquoteright s terms, the LP problem can be solved by searching for a virtually optimal solution that satisfies the objective function. The objective in this situation is to find an interior point in the area defined by all other hyperplanes, except the $A_k X$ hyperplane, which has the highest value of $A_k X$. In another word, the solution should be the furthest point in the direction of the $A_k X$ hyperplane. If the LP can find an optimal solution that is larger than $b_k$, which means the interior point of the area is above the covered region of the inequality $A_k X \leq b_k$. In this case, the hyperplane $A_k X \leq b_k$ intersects with the polytope above, and therefore it is essential.

\begin{algorithm}[!t]
\scriptsize
\SetKwInOut{Input}{Input}
\SetKwInOut{Output}{Output}
\Input{Set of inequations $\{A_jX \leq b_j\ |\forall j \in J\}$}
\Output{Set of indices $N$ of essential inequalities}
\SetKwInOut{Output}{Initialize}
\Output{$N:= \emptyset$,\\
an interior point $x_0$}

\While{$J \neq \emptyset$}
{
	Select an index $k$ from $J$.\\
	Test whether $A_k X \leq b_k$ is \textit{redundant} by solving $LP(N \cup k,k)$ with optimal solution $X^*$. \\
	\uIf{$essential$}
	{
		Produce a ray $(X^*-x_0)$.\\
		Find intersection $z_k$ between $A_k X = b_k$ and $(X^*-x_0)$\\
		\For{$j \neq k$}
		{
		    Find intersection $z_j$ between $A_j X = b_j$ and $(X^*-x)$ \\
		    \If{$||z_j-x_0|| < ||z_k-x_0||$}
		    {
		        $found\_an\_essential\_index = True$.\\
		        \textit{break}.
		    }
		}
	}
	\Else
	{
		$found\_an\_essential\_index = False$.
	}

	\If{$found\_an\_essential\_index$}
	{
		$N:= N \cup \{j\}$\\
		$J:= J \setminus \{j\}$
	}
}

\textbf{return} $N$

\caption{Hyperplane redundancy removal with Clarkson’s algorithm (H-redundancy removal).}\label{Alg:redundancy_removal}
\end{algorithm}

The overall procedure of determining the overlapping areas and generating rule representations from neural networks are introduced in Algorithm~\ref{Alg:knowledge-sharing-NNs}.

\begin{algorithm}[!htb]
\footnotesize
\SetKwInOut{Input}{Input}
\SetKwInOut{Output}{Output}
\Input{Sender and receiver networks: $\mathcal{N}_S$ and $\mathcal{N}_R$; \\
Sender rule set: $\mathcal{P}_S=\{P_{S_1},P_{S_2},...,P_{S_M}\}$; \\
Receiver rule set: $\mathcal{P}_R=\{P_{R_1},P_{R_2},...,P_{R_N}\}$; \\
Performance metrics function: $\Upsilon(\cdot)$}
\Output{New receiver rule set $\mathcal{P}_R'$}

Initialize receiver rule set $\mathcal{P}_R' := \emptyset$. \\
Sample a set of evaluation data $\Phi(X)$ from the target problem. \\
Sort every $X \in \Phi(X)$ into subsets of interior points $\chi_{R_k}$ and $\chi_{S_j}$ of polytopes $P_{R_k} \in \mathcal{P}_R$ and $P_{S_j} \in \mathcal{P}_S$ respectively. \\

\For{polytope $P_{R_k} \in \mathcal{P}_R$}
{
	\For{polytope $P_{S_j} \in \mathcal{P}_S$}
	{
	   \If{$\Upsilon(\mathcal{N}_S(\chi_{S_j})) > \Upsilon(\mathcal{N}_R(\chi_{R_k}))$}
        {
    		$intersection=check\_intersection (P_{R_k},P_{S_j})$ (Algorithm~\ref{Alg:check_intersection}). \\
    		\uIf{$intersection = True$}
    		{
    			Find overlapping polytope $P_{R_k \cap S_j}$ by removing redundant hyperplanes/inequations from a set of all hyperplanes/inequaitions in $P_{R_k}$ and $P_{S_j}$ (Algorithm~\ref{Alg:redundancy_removal}).\\
    			Add $P_{R_k \cap S_j}$ into $\mathcal{P}_R'$.\\
    		}
    		\Else
    		{
    		    Add $P_{R_k}$ into $\mathcal{P}_R'$.\\
    		}
	    }
	}
}

\textbf{return} $\mathcal{P}_R'$.
\caption{Knowledge Integration from Sender and Receiver\textquoteright s Rule Sets}\label{Alg:knowledge-sharing-NNs}
\end{algorithm}

\subsection{Representation Back-Conversion}\label{jpaper2-methodology-fusion-integration}

\begin{figure}[!htb]
\centering
    \subfloat[][Type-1 Network]{\includegraphics[width=0.35\linewidth]{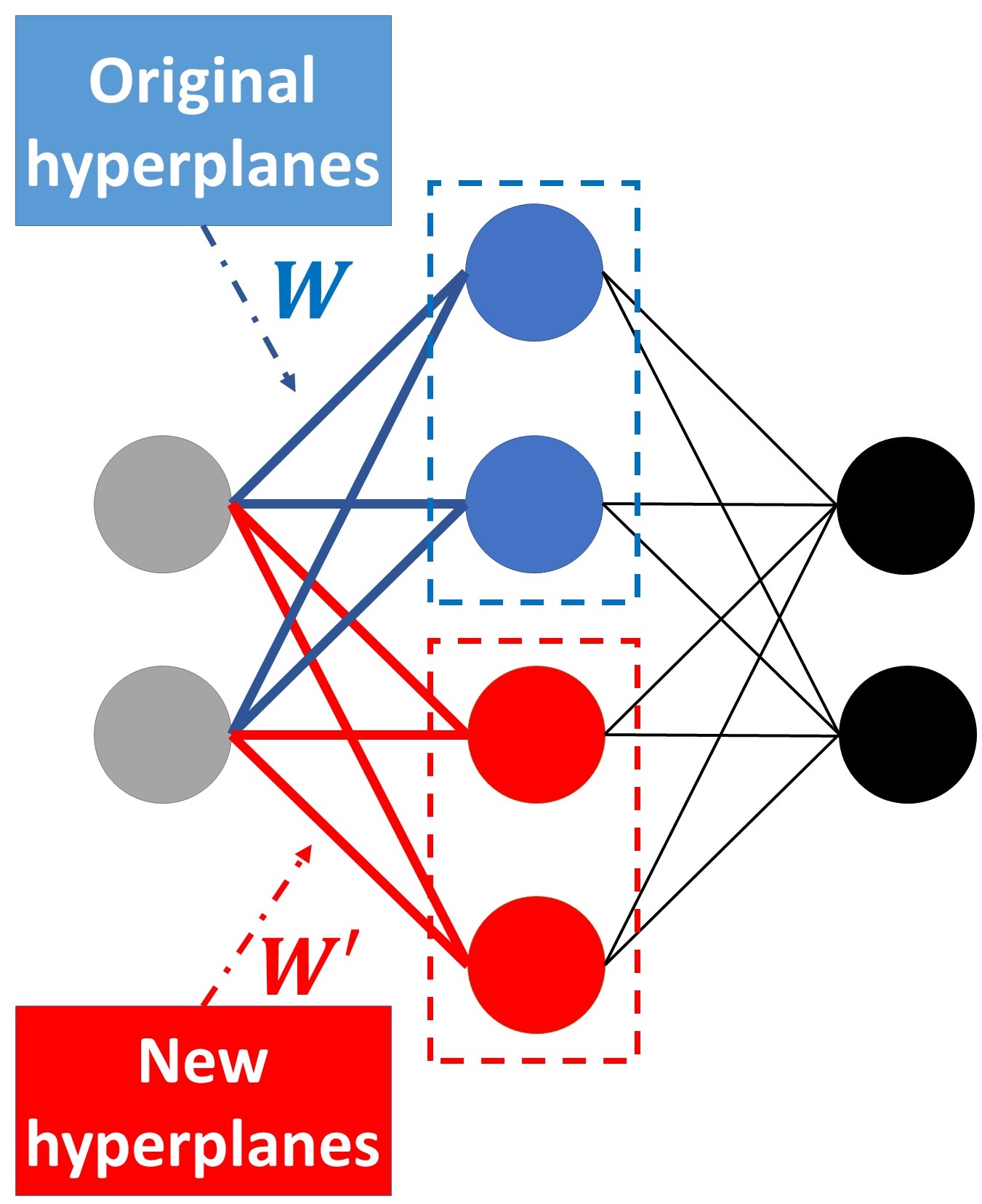}\label{fig:NN-type1}}%
    \hspace{1em}
    \subfloat[][Type-2 Network]{\includegraphics[width=0.45\linewidth]{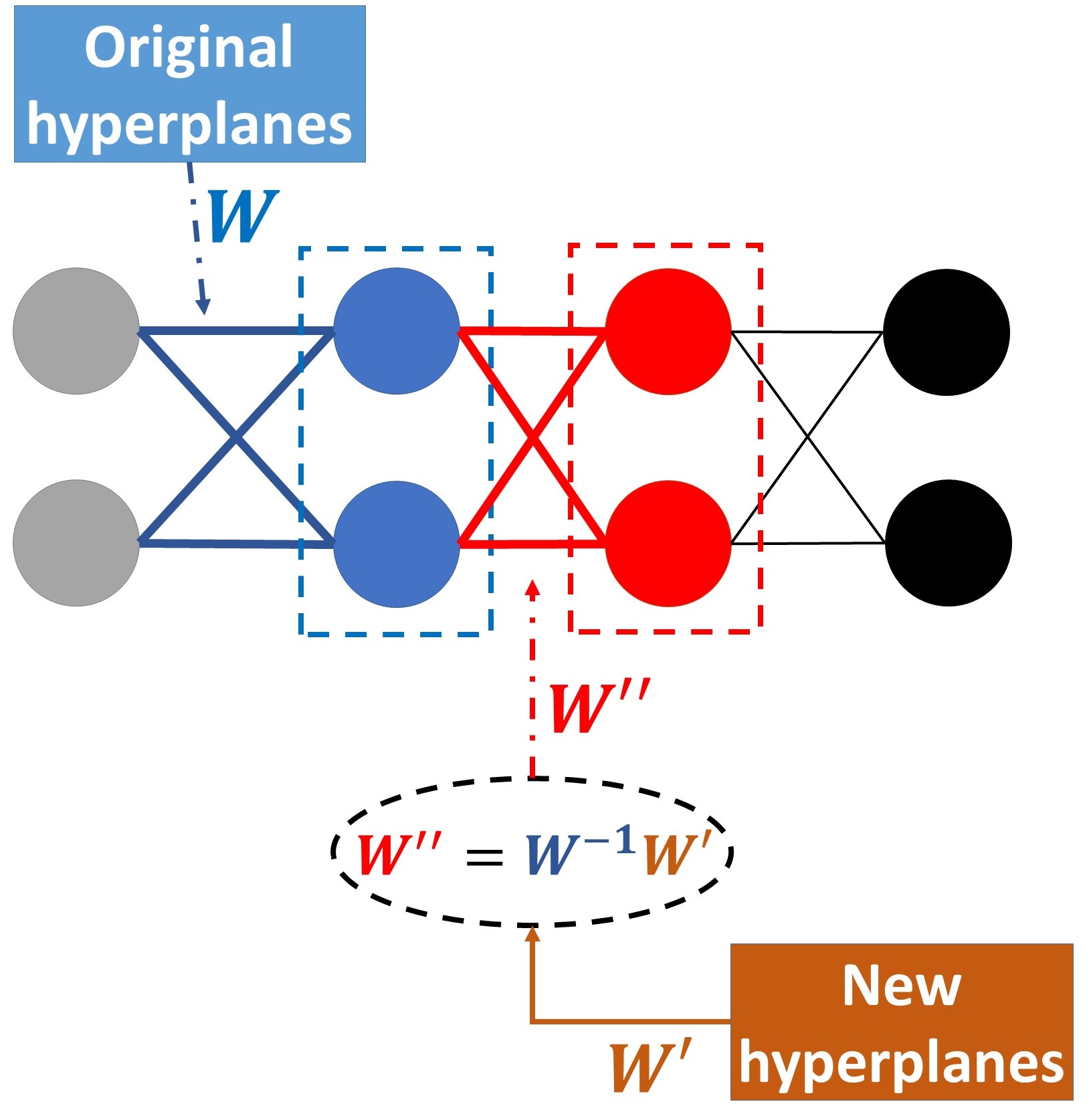}\label{fig:NN-type2}}
    \caption{Two types of neural networks converted from the integrated rule set.}%
\label{fig:two-types-NN}
\end{figure}

The regions in the state space represented by the set of linear inequalities in the rules can be converted into a neural network representation. The weights of the neurons in the network should reflect exactly the hyperplanes determined by the ruleset. The new neural network of the receiver agent after knowledge fusion can be determined by two types of network structure.

The type-1 neural network has the same number of input and output nodes, and only one hidden layer. Each node in the hidden layer represents one hyperplane which is specified by the constraints in the rule. Hence, a specific configuration of activations of the nodes in this layer corresponds to a specific polytope/region, specified by an equivalent rule, in the state space. Figure~\ref{fig:NN-type1} illustrates the structure of the type-1 neural network. The weights and biases of the inequalities in the rules will be assigned to the weights of the links connecting the inputs to the hidden nodes and the biases of those nodes respectively. These weights and biases serve as the kernel initialization for the neural network before the re-training process.
The size of the type-1 neural network increases linearly with the increase of the number of hyperplanes represented by the integrated ruleset. However, the type-1 NN can represent precisely the fused knowledge as it has high comprehensiveness.

Type-2 neural network preserves the same number of input and output nodes, but the new pieces of knowledge are represented by adding a new hidden layer into the neural network. Figure~\ref{fig:NN-type2} illustrates the structure of the type-2 neural network. The weights and biases of the receiver\textquoteright s original inequalities will be assigned to the weights of the links connecting the inputs to the first hidden layer and those nodes\textquoteright \ biases, denoted $W$ and $B$ respectively. Let the weights and biases of the receiver\textquoteright s new inequalities denoted as $W'$ and $B'$ respectively. Note that the weights of the new inequalities specify the relationship between the inputs and the hyperplanes. Therefore, the weights of the connections between the first and second hidden layers can be computed by the following matrix multiplication:
\begin{equation}
    W'' = W^{-1}W'
\end{equation}
The biases of the nodes in the second hidden layer are then:
\begin{equation}
    B'' = B'-BW''
\end{equation}
These weights and biases serve as the kernel initialization for the neural network before the re-training process. The type-2 NN increases the depth of the neural network. The benefit of this type of network is more compact and can represent more hyperplanes with the same number of nodes in the hidden layer compared to the type-1 neural network. However, it needs to sacrifice the representative power and interpretability of the network.

In the next section, we describe a synthetic problem through which we test the performance of two types of neural networks converted from rules after knowledge integration to select the more appropriate method for the framework.

\subsection{Retraining Receiver Agent in a New Environment}\label{jpaper2-methodology-fusion-retraining}

The proposed framework maintains a dual representation system, including the new neural network structure, as a product of back-conversion, as a prediction model and the interpretable set of rules as a reasoner. Based on the performance of the agent after knowledge fusion, we can decide whether to retrain it to adapt to the new problem space or not. This evaluation can detect how sub-spaces represented by the rules contribute to mission successes and failures. The use of rule representation might be transformed into a visual explanation, which provides adequate transparency to help improve the process of identifying the areas of the input space where the model performs suboptimally.

In the scope of this paper, we use reinforcement learning to train agents. We propose a retraining technique, called retraining with Priority on Weak State Areas (PoWSA). This technique is made possible as our proposed framework enables the interpretable representation of knowledge in the form of rule sets. This algorithm assesses whether decision polytopes — specified by decision rules, each covers a different area in the problem space, containing input states which contribute directly to the failure of missions. Such kind of polytopes/rules is called weak polytopes/rules. In this paper, we use Q-learning with experience replay to update neural networks. As the Q-learning algorithm estimates the value function of a state based on temporal difference (TD) learning, the states which are close to weak polytopes have a higher probability to transit to the weak polytopes, leading to a higher indirect contribution to failure. Therefore, our modified algorithm assigns a lower priority to data which having a higher distance to the centroid of the weak polytope, to be used in retraining. Hence, the updated priority of sampling a data sample from the replay buffer is computed by:
\begin{equation}
\rho '(X) = e^{(\frac{-0.25d}{\sqrt{2}L-d_{min}})}\rho (X)
\end{equation}
where $L$ is the size of the environment, $d$ is the Euclidean distance between the data point and the centroid of the closest weak polytope, $d_{min}$ is the distance from the centroid of the closest weak polytope within which the priority is always $1$. $\rho (X)$ is the priority of data $X$ in the original prioritized experience replay, which is replaced by $\rho '(X)$ in our method. Similarly, the updated exploration rate of the agent is computed by:
\begin{equation}
{\varepsilon '}_t = e^{(\frac{-0.5D}{\sqrt{2}L-d_{min}})}\varepsilon_t
\end{equation}
where $D$ is the Euclidean distance between the centroid of the polytope containing the current state of the agent and the centroid of the closest weak polytope.
The evaluation metrics of the retraining methods are the task \textit{success rate}, and the \textit{number of steps of successful missions}.

In layman\textquoteright s terms, in our proposed retraining method, a higher priority is placed on sampling previous states belonging to sub-spaces associated with failures. The exploration-exploitation rate of interactions with environments in the retraining process is also non-uniform, with an exploration factor computed based on the distance between the subspace of interest to sub-spaces associated with failures. This modification of the learning algorithm is expected to promote retraining in sub-spaces associated with weak chunks of an agent\textquoteright s knowledge while reducing the cost of retraining for strong areas.

\section{Experiments}\label{jpaper2-experiments}

This section first describes two problems we use for demonstrating our proposed interpretable knowledge fusion framework. The main problem, between the two, is the autonomous swarm guidance problem which is described in detail along with the simulation environment and training procedure. Then evaluation metrics are introduced for assessing the effectiveness of our knowledge fusion framework, followed by an in-depth discussion of each experiment. The code for reproducing our results in this paper is available from the GitHub repository: \url{https://github.com/tudngn/IKTF-NN}.

\subsection{Problem Spaces}
\subsubsection{Synthetic Binary Classification}\label{jpaper2-experiments-integration}
We test the performance of two types of neural networks on the following synthetic data sets. The receiver agent is originally trained on the binary classification task P1:
\begin{equation}
f = \begin{cases}
        0, & \mbox{for} \;\; (2x+y+2<0) \land (x+2y+3<0) \\
        1, & \mbox{otherwise}
    \end{cases}
\label{eq:P1}
\end{equation}

The sender agent is originally trained on the binary classification task P2:
\begin{equation}
f = \begin{cases}
        1, & \mbox{for} \;\; (4x+5y-8<0) \land (3+4y-9<0) \\
        0, & \mbox{otherwise}
    \end{cases}
\label{eq:P2}
\end{equation}

The receiver agent receives the knowledge from the sender agent so that its new knowledge structure can solve the binary classification problem P3:
\begin{equation}
f = \begin{cases}
        1, & \mbox{for} \;\; (2x+1y+2<0) \land (1+2y+3\geq 0) \land
        \\
        & \;\;\;\;\;\; (3+4y-9\geq 0)\\
        1, & \mbox{for} \;\; (2x+1y+2\geq0) \land (1+2y+3<0) \land
        \\
        & \;\;\;\;\;\; (4+3y-8<0)\\
        1, & \mbox{for} \;\; (2x+1y+2\geq0) \land (1+2y+3\geq0) \land \\
        & \;\;\;\;\;\; (4+3y-8<0) \land (3+4y-9<0)\\
        0, & \mbox{otherwise}
    \end{cases}
\label{eq:P3}
\end{equation}

\begin{figure*}[!htb]
\centering
    \subfloat[][P1]{\includegraphics[width=0.28\linewidth]{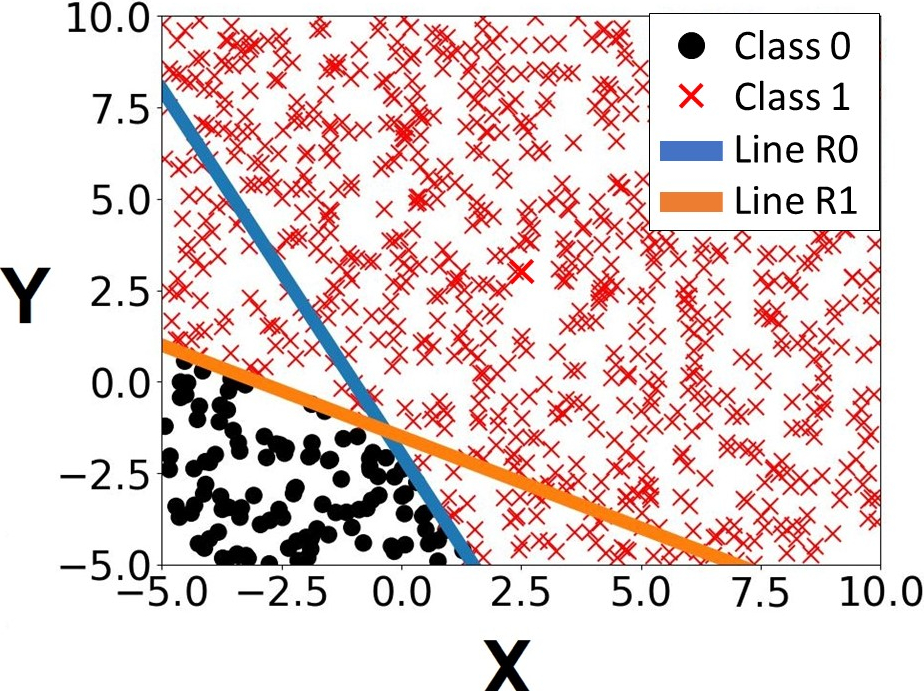}}%
    \hspace{1em}
    \subfloat[][P2]{\includegraphics[width=0.28\linewidth]{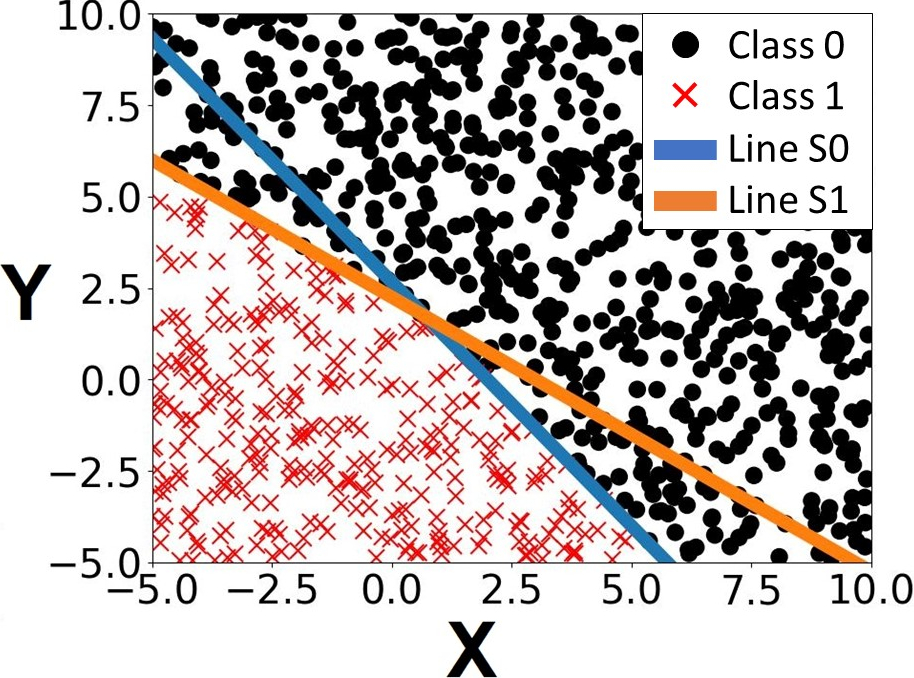}}%
    \hspace{1em}
    \subfloat[][P3]{\includegraphics[width=0.28\linewidth]{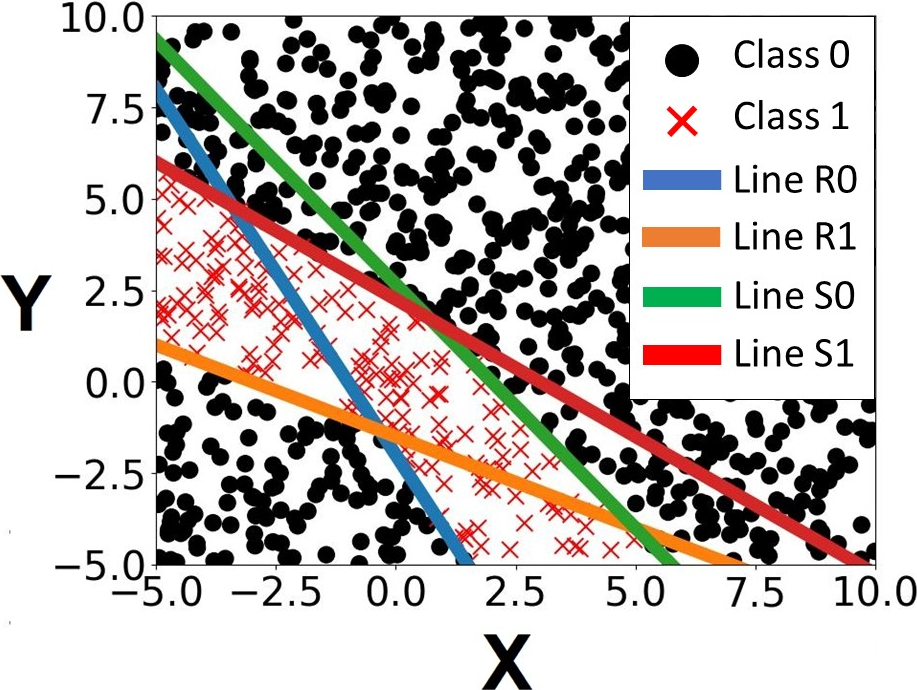}}%
    \caption{The distribution of data samples in three synthetic problems.}
\label{fig:three-problems}
\end{figure*}

Figure~\ref{fig:three-problems} demonstrates the distribution of classes of data samples in the input space for all problems. These problems are simple classification problems that can be served as a proof-of-concept for demonstrating the feasibility of two types of conversion techniques toward a more complex problem.

The agents are originally trained on P1 and P2 problems with 5000 data samples. Each agent uses a simple neural network with two input nodes ($X$ and $Y$), a single hidden layer with two nodes, and a single output with a sigmoid function. Binary cross-entropy is used as the loss function. Theoretically, two hidden nodes are enough to estimate the two hyperplanes in the P1 and P2 problems. For addressing the problem P3 with four hyperplanes, the receiver then fuses the knowledge of its own and the knowledge sent from the sender agent. Two types of neural networks are generated as described in Section~\ref{jpaper2-methodology-fusion-integration}. The networks are then retrained and tested on the new problem P3. This process is run ten times. We compare the performance of the two networks in two cases: the networks are initialized with and without the weights acquired from the ruleset. The latter case is equivalent to training the network from scratch, thus it is the baseline method for training networks without knowledge fusion.

\subsubsection{Shepherding-Based Autonomous Swarm Guidance}\label{jpaper2-experiments-transfer}

The second test will be conducted using the shepherding problem-a complex autonomous control problem. The shepherding model explained in Section~\ref{jpaper2-appendix-shepherding} (Supplementary document) is a reactive swarm-robotic model where the shepherding agent acts based on a weighted sum of all forces exerted by other mobile agents (sheep) and objects in the environment~\cite{El-Fiqilimits2020}. The agent switches between two types of behaviours called collecting and driving, each produces a sub-goal location that attracts the agent. The sub-goal is generated based on predefined rules. For simplification, in this paper, we assume the agent is already skilled with the collecting behaviour. The agent needs to learn the driving behaviour for path planning, which safely herds the flock of sheep towards a target position, by using a deep reinforcement learning approach.

The input states that is observed by the shepherd is a vector of 4 dimensions: (1) Distance between sheep\textquoteright \ GCM and the shepherd, (2) Direction of sheep\textquoteright \ GCM relative to shepherd, (3) Distance between target and sheep\textquoteright \ GCM, and (4) Direction of the target relative to sheep\textquoteright \ GCM. The shepherd can select five possible driving points to move to so that it can drive the sheep to five corresponding  directions (\textit{north}, \textit{northwest}, \textit{west}, \textit{southwest}, \textit{south}).

\begin{figure*}[!tb]

\centering
    \subfloat[][Environment A (least complex)]{\includegraphics[width=0.25\linewidth]{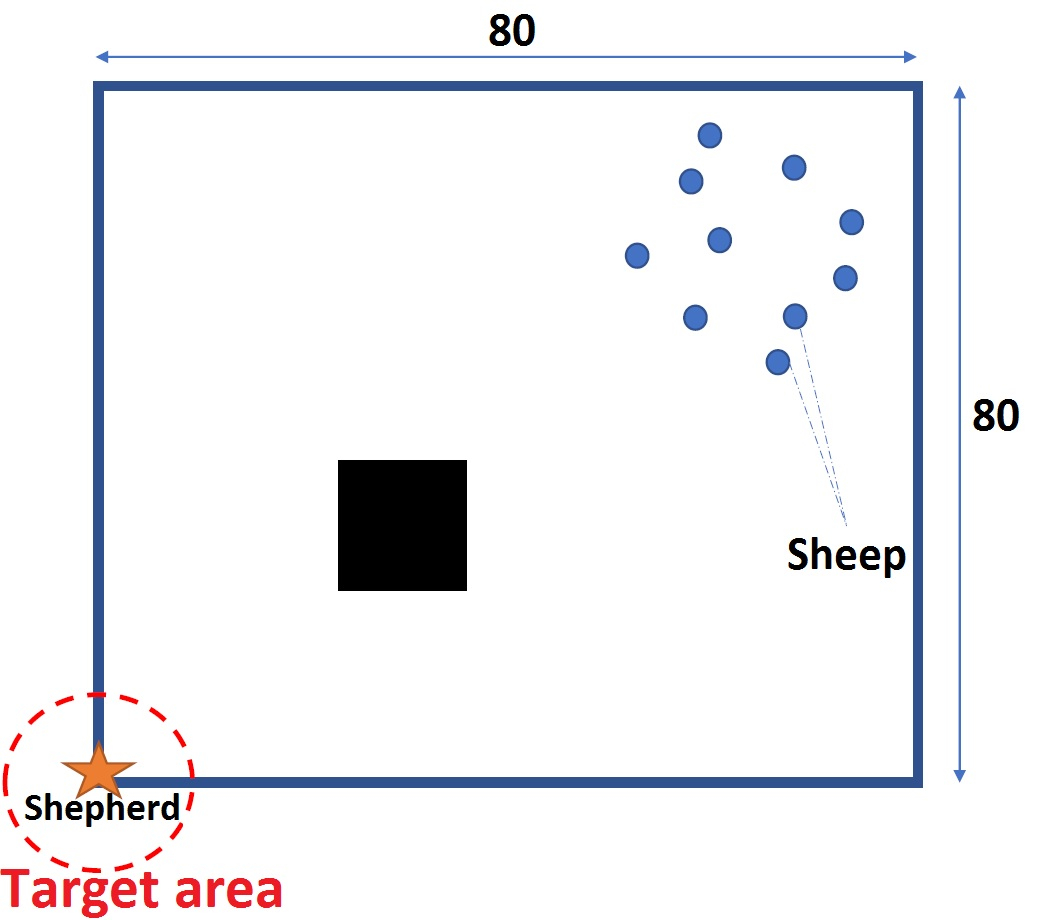}}%
    \hspace{1.5em}
    \subfloat[][Environment B]{\includegraphics[width=0.25\linewidth]{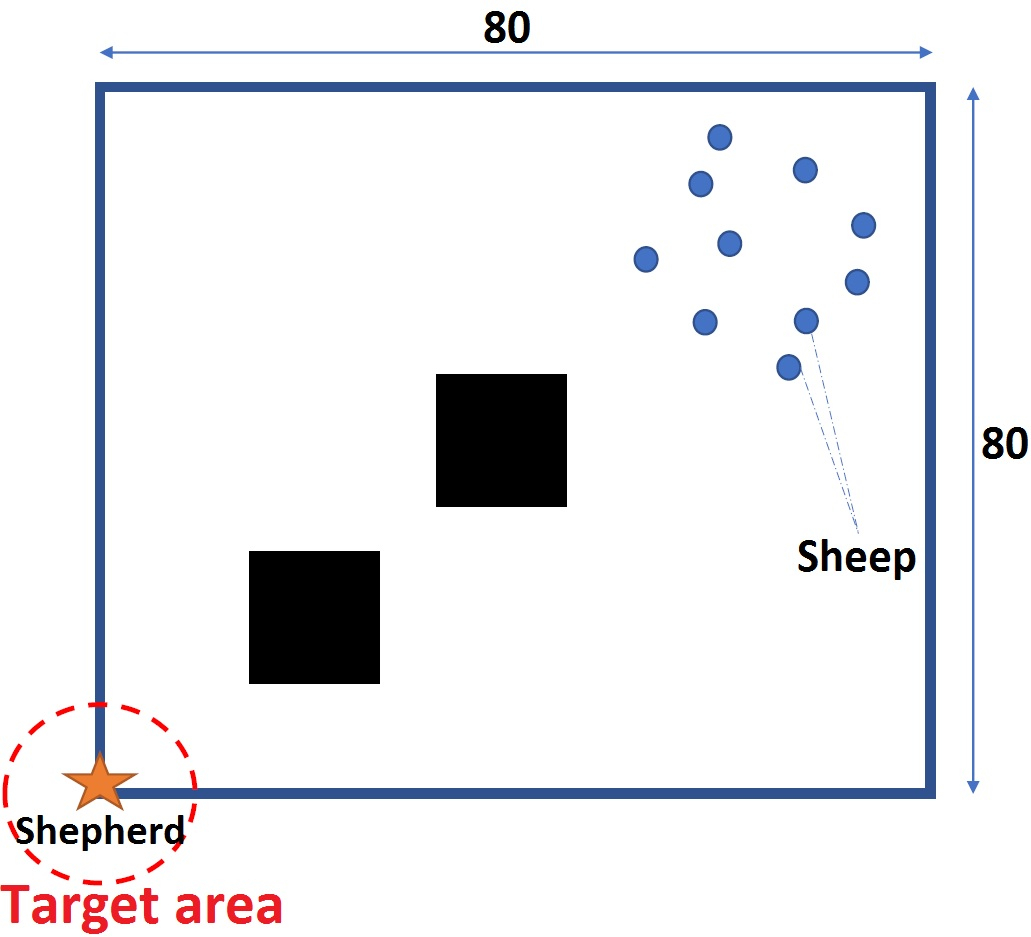}}%
    \hspace{1.5em}
    \subfloat[][Environment C (most complex)]{\includegraphics[width=0.25\linewidth]{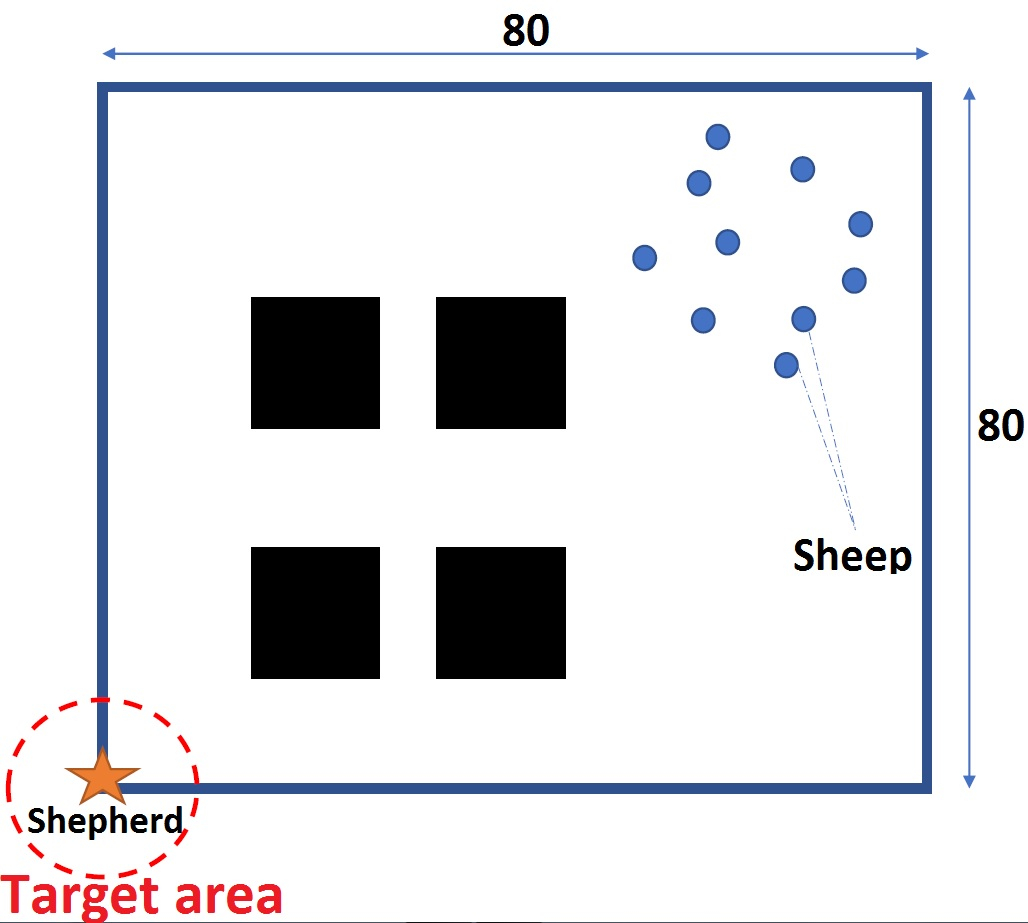}}%
    \caption{The environments are designed to have progressively increasing complexity.}
\label{fig:environments}
\end{figure*}

Three environments are designed as demonstrated in Figure~\ref{fig:environments}. The receiver and sender agents are trained in environments A and B respectively while environment C is the new environment to test the performance of the knowledge fusion framework. The complexity of the learning process required to obtain enough knowledge to achieve satisfactory performance of the tasks increases from environment A to environment C.

The reinforcement learning algorithm used in this paper is Double Deep Q-Network (DDQN) with prioritized experience-replay (PER). The algorithm uses a neural network to approximate state-action values and be able to generalize for a large continuous state space. A reinforcement learner\textquoteright s experiences of transitions are stored and randomly sampled to perform a batch training process for better data efficiency and to eliminate the effects of correlation in the dataset due to temporal dependencies. However, it might be that the agent will fail to produce a path that avoids the risk of collisions between the sheep and the obstacle or has a low chance of reaching the target in early training sessions. In that case, the imbalance between the bad and good experiences is immense, leading to a more biased data sampling. Therefore, we wish to use a prioritized experience-replay buffer with a higher probability of sampling rare, but significant, experiences. The training algorithm and parameters are described in Section~\ref{jpaper2-appendix-NNtraining} (Supplementary document).

After training sender and receiver\textquoteright s networks, the knowledge of each agent is interpreted into decision rules and the fusion process is implemented, as explained in Section~\ref{jpaper2-methodology}, is performed.
The retraining stage is performed ten times with random seeds. The proposed method (knowledge fusion with and without PoWSA) is compared with three baseline methods including training NN models (1) from scratch, (2) with A2T architecture, and (3) MULTIPOLAR architecture. For models trained from scratch, the architecture of NN is similar to the one after our proposed knowledge fusion method. For A2T framework, it uses three-block architecture including source NNs that provide pre-trained knowledge, a base model that learns new knowledge and an attention network that fuses the outputs of source and base models~\cite{rajendran2015attend}. In this paper, the sender and receiver networks are assigned as the source models in A2T, while the base and attention models are initialized with the same structure as the sender and receiver networks, but with small random weights. For the MULTIPOLAR framework, it uses a three-block model including source NNs, an auxiliary network that estimates residuals, and a trainable aggregation weight matrix to fuse the outputs of the networks~\cite{barekatain2020multipolar}. Similarly, the source NNs are the sender and receiver networks while the auxiliary network, which has the same structure as other networks, is initialized with small random weights. The aggregation weight matrix is initialized with the all-ones matrix as suggested by~\cite{barekatain2020multipolar}.     

The performance of the models generated by our proposed knowledge fusion framework is evaluated through 1000 tests on the new environment C.

\subsection{Evaluation Metrics}

In this section, we introduce three evaluation metrics for the knowledge fusion framework as follows:

\begin{itemize}
    \item \textbf{Task performance}: In experiment 1, we compare the performance of two types of NNs that are converted back from the rule representations. The metrics for this experiment is the classification \textit{accuracy}. For the main problem (shepherding), the evaluation metrics include the \textit{task success rate} and \textit{the number of steps} required to complete the missions successfully. The number of steps in fail test runs will be excluded from the computation of the latter metrics. 
    \item \textbf{Training-sample efficiency}: We investigate the behaviour of the learning curves and determine the required time for training until convergence. For experiment 1, the curves of loss values during training are examined while the reward-per-step curves are examined for experiment 2.
    \item \textbf{Transparency}: The decision-boundaries created by rule representations are investigated. This is accomplished by visualising the hyperplanes and decision polytopes in the state space that match the constraints constituting the rules. The sets of rules from receiver and sender agents and the integrated set of rules after knowledge integration are visualized so that we can have some insights into how the fusion process works.
\end{itemize}

\subsection{Results and Discussion}\label{jpaper2-results}

In this section, we evaluate the performance of the proposed approach. Firstly, we investigate the performance of two different knowledge representation back-conversion techniques into a new NN structure on simple synthesis classification problems. Secondly, we evaluate the performance of models after knowledge fusion against baseline training methods in the shepherding problem. The results were generated on an Intel Core i7-7700 CPU and NVIDIA GeForce GTX1080 GPU.

\subsubsection{Neural Network Back-Conversion}\label{jpaper2-results-integration}

\begin{figure*}[!t]
\centering
    \subfloat[][]{\includegraphics[width=0.23\linewidth]{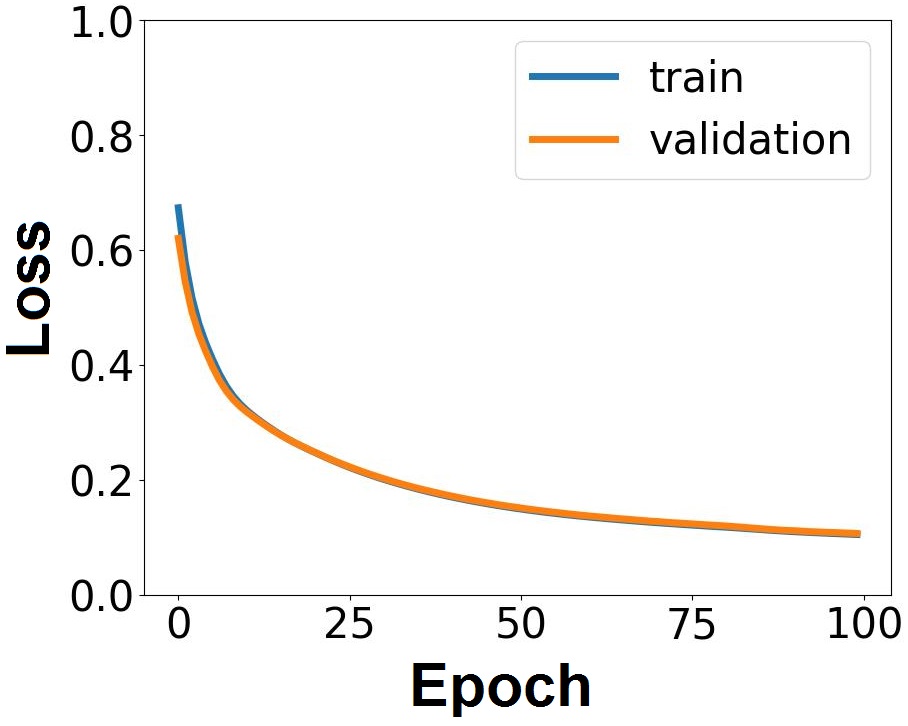}}%
    \hfill
    \subfloat[][]{\includegraphics[width=0.23\linewidth]{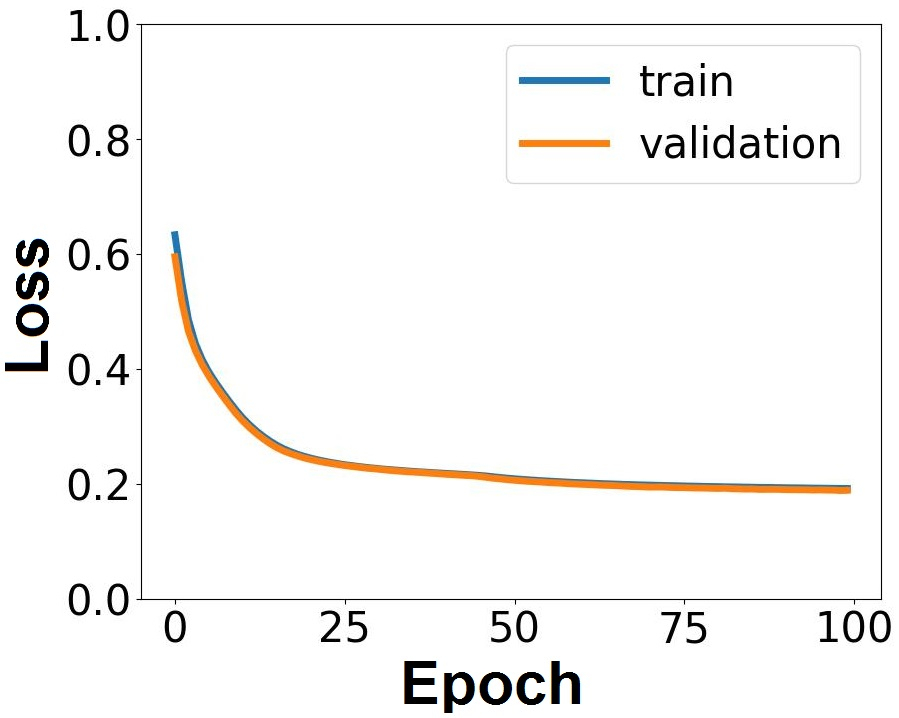}}%
    \hfill
    \subfloat[][]{\includegraphics[width=0.23\linewidth]{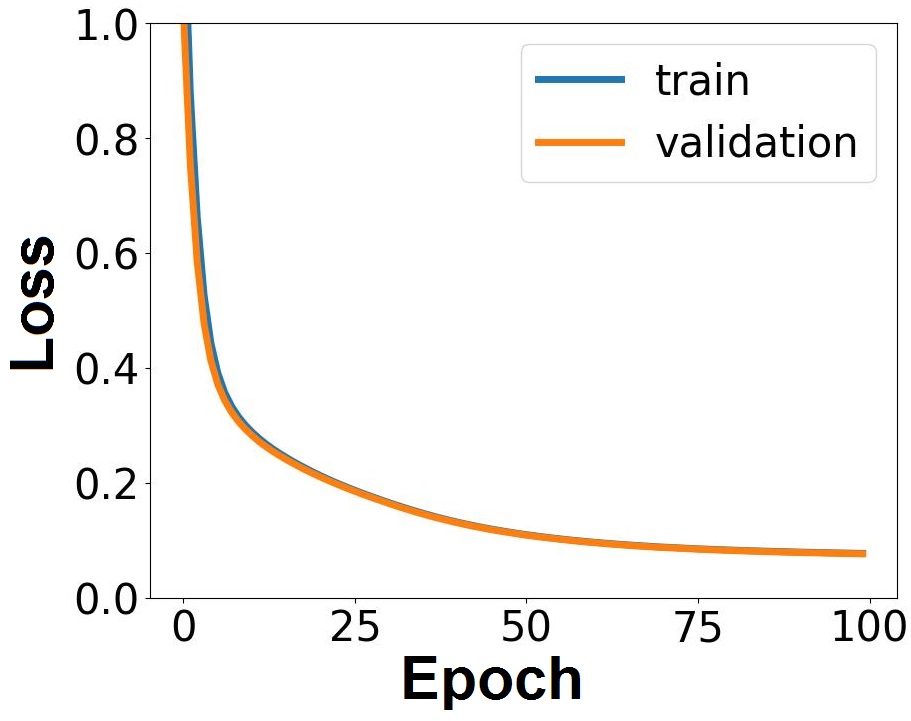}}%
    \hfill
    \subfloat[][]{\includegraphics[width=0.23\linewidth]{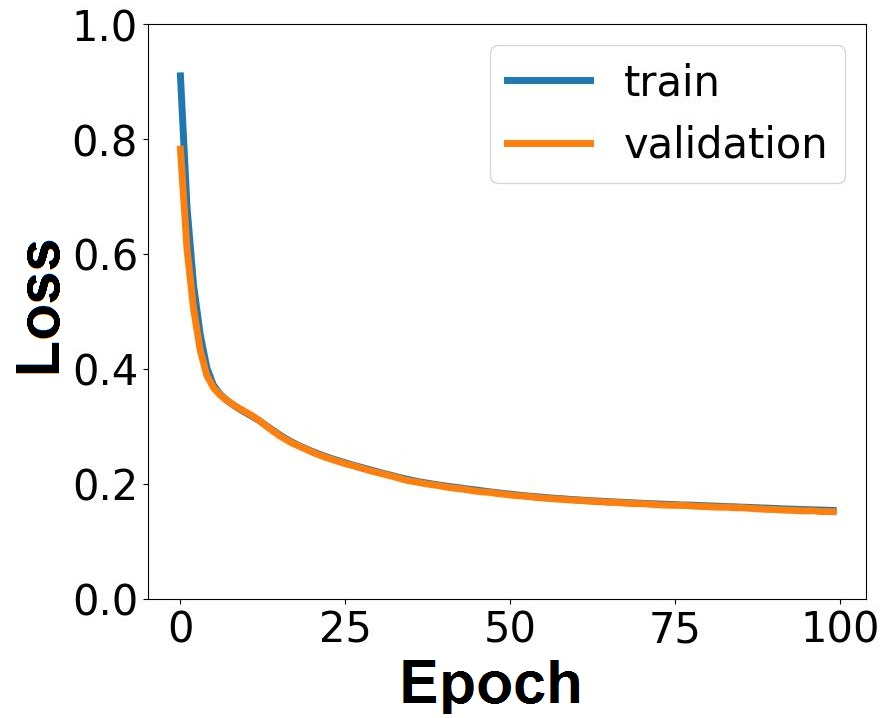}}%

    \caption{Learning curves in four cases: (a) Type-1 NN - trained from scratch, (b) Type-2 NN - trained from scratch, (c) Type-1 NN - retrained after knowledge fusion, and (d) Type-2 NN - retrained after knowledge fusion.}
\label{fig:P3-losscurve}
\end{figure*}

The learning curves as a result of the training processes with type-1 NN and type-2 NN are shown in Figure~\ref{fig:P3-losscurve}. For type-1 NN, the model with fused knowledge has a faster reduction of the loss over training epochs than the model of the same architecture but trained from scratch. Fusing the knowledge for this type of model helps achieve a loss of around 0.15 after only less than 40 training epochs while it takes approximately 100 training epochs for the model trained from scratch to lower the loss to a similar value. This indicates that training the model with the type-1 conversion of fused knowledge accelerates the learning process. On the other hand, the loss after training a type-2 NN model with knowledge fusion is also lower than that of a model without knowledge fusion. The loss values of these models are higher than 0.2 at the end of 100 epochs, which are much higher than the type-1 counterparts.

When testing the NNs on newly generated data, integrating knowledge to the NN of a receiver with the type-1 technique increases the average classification accuracy than training the model from scratch. The training process is also more stable, demonstrated by a lower standard deviation. The performance of type-1 models is $5$-$6\%$ higher than that of the type-2 models.

Although there is no significant overfitting of the models with type-2 architecture as suggested by the similar learning curves of training and validation processes, the type-2 architecture seems to over-complicate the decision outputs by introducing too many hyperplanes than necessary. The maximum number of hyperplanes represented by the type-1 model is $(N + M)$ while the number for the type-2 model is $(N + 2^N\times M)$, where $N$ and $M$ are the numbers of receiver\textquoteright s original hyperplanes and the number of new hyperplanes received from sender respectively. In a training process with a backpropagation algorithm, the errors propagated back to the hidden nodes are more independent of one another than the errors in the case of deeper networks like the type-2 model. The hyperplanes in later layers of deeper models are also the linear combinations of values from the previous layer, thus they are more conditionally dependent. These factors might make a type-2 model more susceptible to errors.

As the type-1 conversion technique achieves higher performance, we select it to use in the knowledge fusion framework. This type of network also has less layer of transformation, which may benefit the interpretation process if required in further communication between agents.

\begin{table}[!b]
\vspace{-1em}
\centering
\small\addtolength{\tabcolsep}{-3pt}
\caption{The prediction accuracies of the models on the test set.}
\label{tab:P3-accuracy}
\begin{tabular}{lll}
                                   & \textbf{Type-1 NN}        & \textbf{Type-2 NN}        \\ \hline
\textbf{Trained from scratch}              & $96.12 \pm 1.07$ & $91.56 \pm 5.69$ \\
\textbf{Trained after knowledge fusion} & $97.09 \pm 0.57$ & $91.84 \pm 7.05$
\end{tabular}
\end{table}

\subsubsection{Sample Efficiency and Task Performance}\label{jpaper2-results-retraining}

\begin{figure*}[!htb]
\centering
    \subfloat[][Training from scratch\label{fig:baseline-training-curve}]{\includegraphics[width=0.28\linewidth]{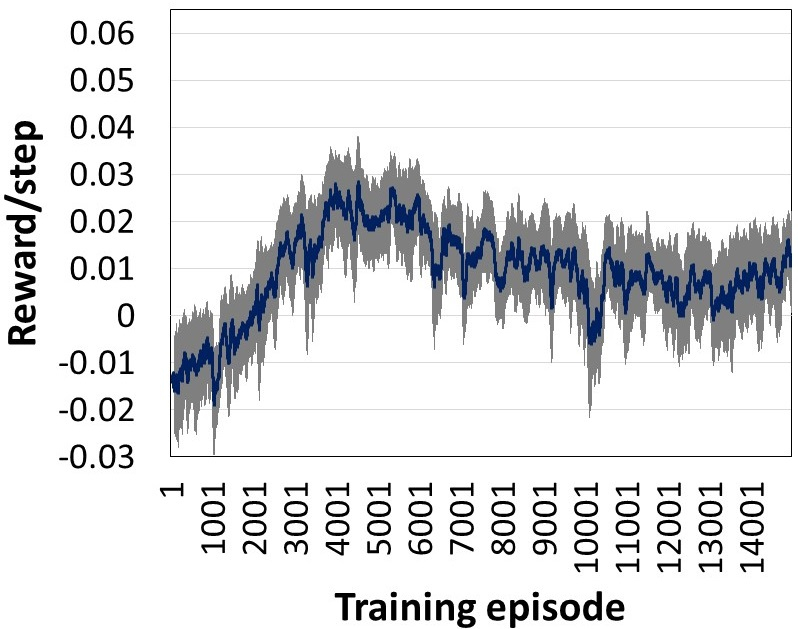}}
    \hspace{1em}
    \subfloat[][A2T\label{fig:A2T-training-curve}]{\includegraphics[width=0.28\linewidth]{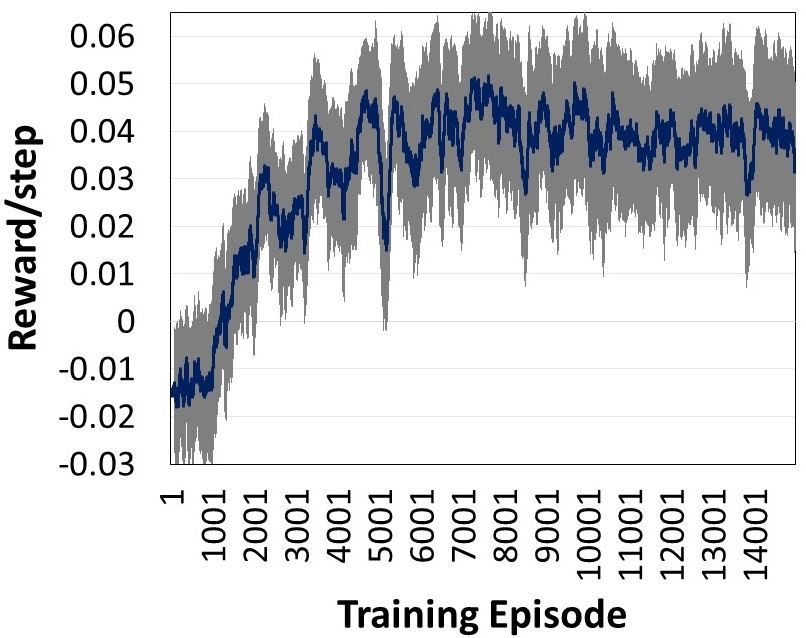}}
    \hspace{1em}
    \subfloat[][MULTIPOLAR\label{fig:MULTIPOLAR-training-curve}]{\includegraphics[width=0.28\linewidth]{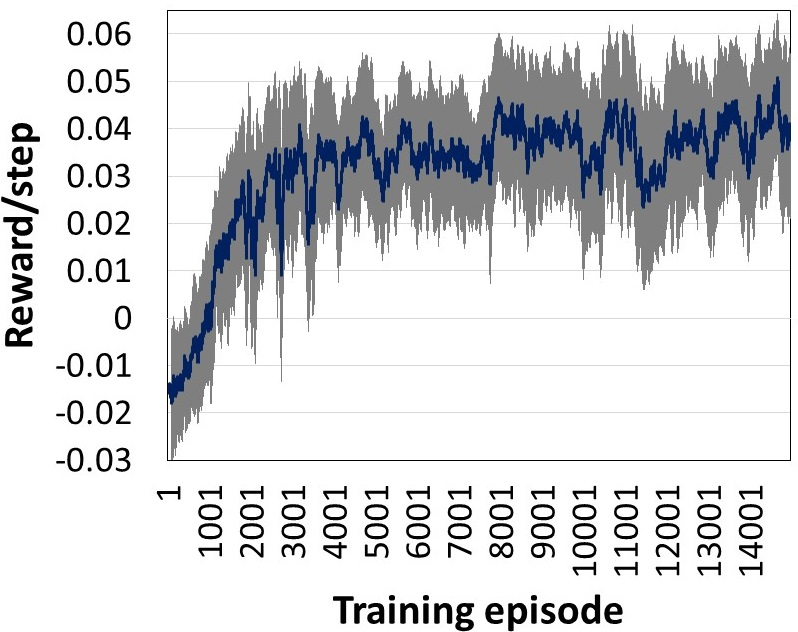}}
    \\
    \subfloat[][IKF\label{fig:training-curve-with-transfer}]{\includegraphics[width=0.28\linewidth]{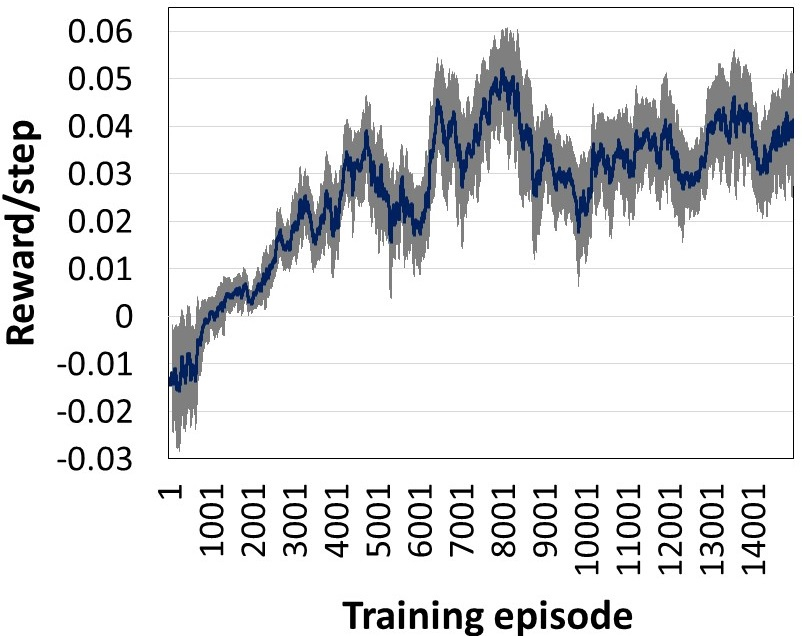}}
    \hspace{1em}
    \subfloat[][IKF-PoWSA\label{fig:training-curve-with-transfer-PowSA}]{\includegraphics[width=0.28\linewidth]{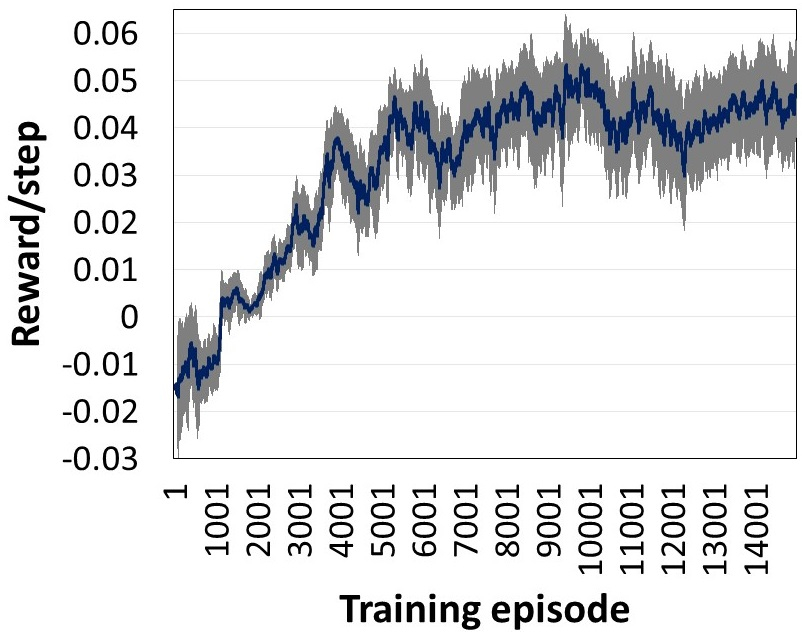}}
    \caption{The training curves of models with different frameworks in Environment C. The blue line indicates the change in the reward received per learning steps as training proceeds. The value of each point in the curve is averaged over ten runs of training. The grey shades indicate the standard deviation at corresponding episodes.}
\label{fig:retraining-curve}
\end{figure*}

Firstly, we compare the training curves of new models generated with our proposed knowledge fusion framework and baseline models. Figure~\ref{fig:baseline-training-curve} shows the average reward per learning step over training episodes of models trained from scratch. At the end of the training session, the models only achieve a reward/step of 0.015 on average. On the other hand, when the models take advantage of the previous knowledge with the other four methods, the reward per step can reach approximately 0.04 near the end of the training session.

We also compare the learning processes of models with our proposed knowledge fusion (without the PoWSA technique (IKF), and with POWSA technique (IKF-POWSA)) (Figures~\ref{fig:training-curve-with-transfer} and~\ref{fig:training-curve-with-transfer-PowSA} respectively) against baseline models trained with A2T and MULTIPOLAR frameworks (Figures~\ref{fig:A2T-training-curve} and~\ref{fig:MULTIPOLAR-training-curve} respectively). In the first 4000 episodes, A2T and MULTIPOLAR methods achieve better sample efficiency with a higher rise in episodic reward compared to our proposed knowledge fusion method. However, at the end of the training phase, A2T and MULTIPOLAR models only achieve approximately 0.04 on average. The fluctuation of the average learning curve is large with a relatively higher standard deviation compared to our proposed method, which signifies the instability of the methods in different random seeds.

Our proposed models without the PoWSA technique (IKF) are also unstable in the middle of the training processes. However, the models with PoWSA (IKF-PoWSA) have a higher priority to learn and a higher exploration rate in weak regions of knowledge represented by the decision polytopes in the state space that strongly connects to the failures in the assessment phase before integrating new knowledge. The results suggest that the non-uniform sampling priority and exploration rate produce more stable learning, demonstrated by a milder fluctuation of the learning curve after the 7000-episode milestone. In addition, the reward per step of models that use the PoWSA technique, at the end of the training session, achieves an average of 0.045, which is higher than the value of models without PoWSA and the other baseline models. It can be concluded that the use of the PoWSA technique stabilizes the training process of models after knowledge fusion and helps them achieve a slightly better training objective value.

Secondly, we compare the performance of the models in test shepherding tasks.  Table~\ref{tab:shepherding-performance} records the average task success rate of ten models generated by each method (effectiveness) and the average number of steps (efficiency) that an agent needs to take to complete the task. Note that we only count the number of steps in a successful mission, in which the sheep are collected and herded successfully to the target position.

\begin{table}[!b]
\vspace{-1em}
\small
\centering
\caption{The success rate and the number of steps in successful runs achieved by models generated by three different methods.}
\label{tab:shepherding-performance}
\begin{tabular}{ccc}
    & \textbf{Success rate} ($\%$) & \textbf{Number of steps} \\ \hline
 \textbf{Trained from scratch}
 & $52.29\pm26.99\%$    & $520.43\pm219.58$  \\
 \textbf{A2T}
 & $74.78\pm15.33\%$    & $265.27\pm61.55$  \\
  \textbf{MULTIPOLAR}
 & $84.16\pm6.63\%$    & $256.70\pm58.16$  \\
 \textbf{IKF}
 & $82.86\pm20.17\%$    & $322.59\pm123.93$  \\
 \textbf{IKF + PoWSA}
 & $\mathbf{95.61\pm4.76\%}$     & $305.20\pm129.72$

\end{tabular}
\end{table}

When testing the models trained from scratch with models trained with four knowledge transfer frameworks, the results show a significantly higher success rate than the latter methods (at the significance level of 0.01). The number of steps that the agents need to take to complete the task on average is also lower when the previous knowledge is used to bootstrap the learning of the networks. These results agree with the implication of the training curves from the methods, which obviously show inferior and slow training processes from models when training from scratch. To train the model successfully from the scratch, the NNs might need a more complex architecture and/or a slower exploration rate decay (more exploration time to overcome possible local optima) with more training episodes. 

When comparing models training with different knowledge fusion methods, we observe that the average success rate of A2T models is the lowest while MULTIPOLAR models and IKF achieve similar success rates. IKF-PoWSA achieves approximately $11\%$ higher success rate than models trained with MULTIPOLAR. A further t-test confirms the difference is significant at the level of 0.05. It is also noticeable that the standard deviation of the success rate in the case of using the IKF and the other two baseline methods is higher than the figures from the IKF-PoWSA models. Combined with the rough fluctuation of the reward per step curve in the training session, it is evident that even though accelerated by the fused knowledge, A2T, MULTIPOLAR and IKF models are still unstable.

The use of our proposed interpretable knowledge fusion framework also helps identify the pieces of knowledge that would be valuable if fused, followed by determining the hyperplanes, converting the hyperplanes\textquoteright \ inequations into new neural substrates, and integrating them into the neural networks with non-redundancy.

Regarding computational time for training models, A2T models include two trainable neural networks (base and attention networks), each is trained with different target values for outputs, while MULTIPOLAR has one trainable network and a small weight matrix. Therefore, the average computational time for the training process of A2T models is nearly 2 times longer than the time for MULTIPOLAR. Our proposed framework IKF-PoWSA is roughly $14.5\%$ more time-consuming than MULTIPOLAR as it considers the current state of agents relative to the weak chunks of knowledge every time step for computing the priority of the collected sample and exploration rate.

In summary, our proposed framework with the PoWSA technique is more stable and can generate models with a better success rate than other baseline methods. Even though the method has more computational complexity than MULTIPOLAR, it produces models that have a higher success rate (effectiveness) and an equivalent number of steps required to complete the task (efficiency).

\subsubsection{Transparency of Knowledge Fusion Process}\label{jpaper2-results-transparency}

\begin{figure*}[!htb]
\centering
    \subfloat[][\label{fig:receiver_polytopes}]{\includegraphics[width=0.38\linewidth]{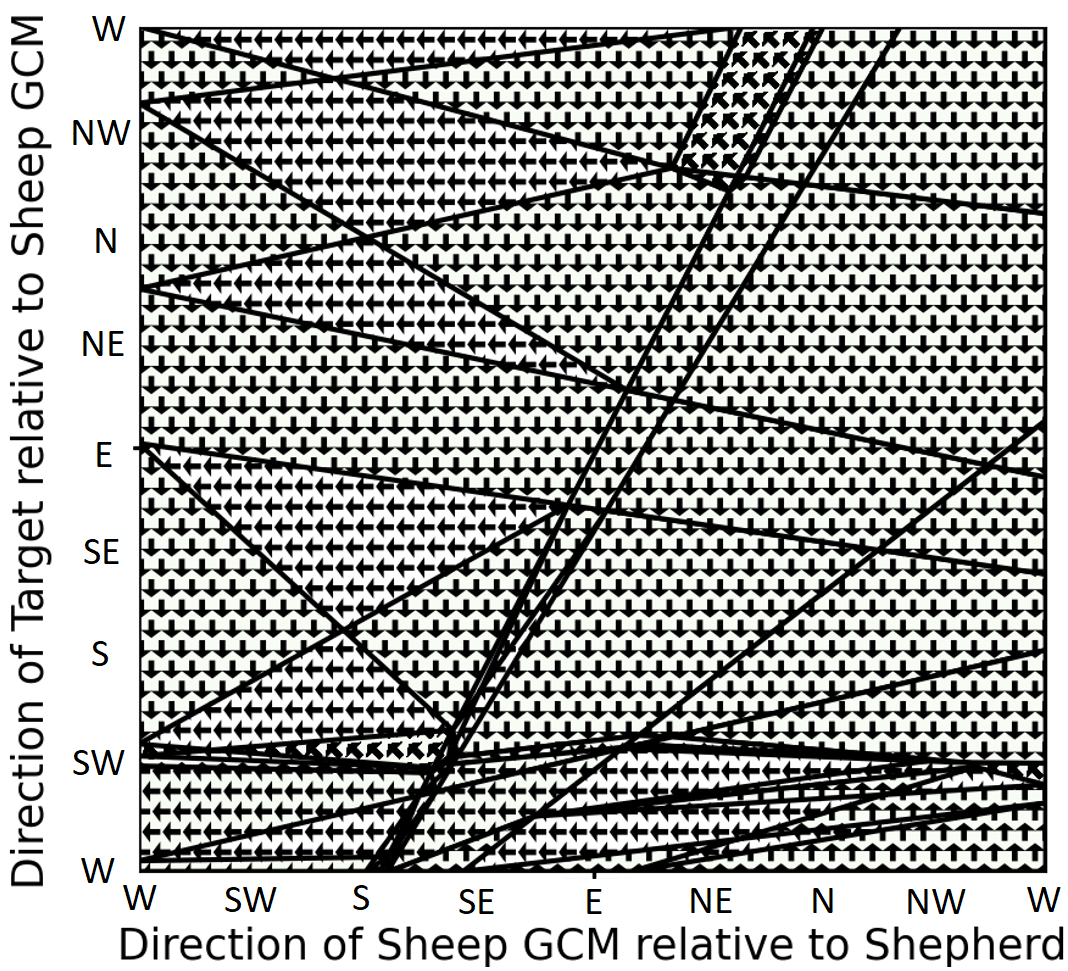}}
    \hspace{1.5em}
    \subfloat[][\label{fig:sender_polytopes}]{\includegraphics[width=0.38\linewidth]{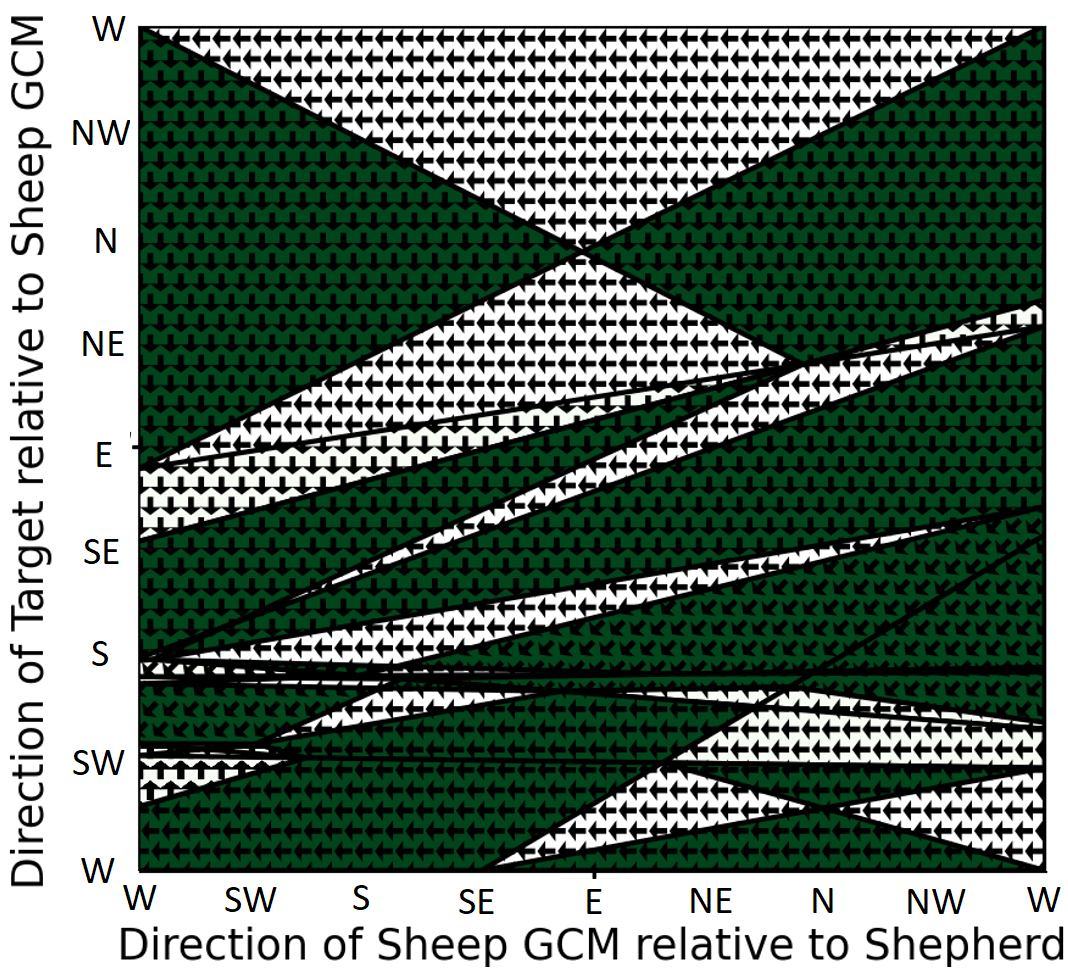}}
    \\
    \vspace{-1em}
    \subfloat{}{\includegraphics[width=0.38\linewidth]{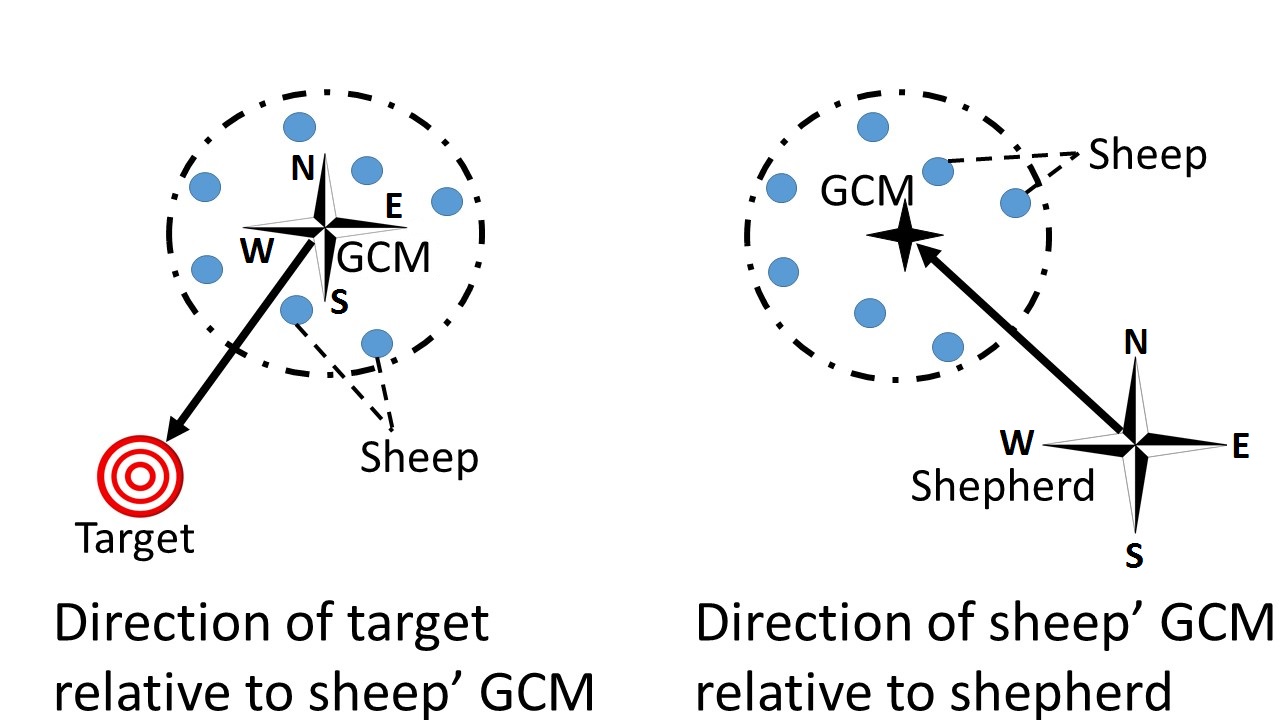}}
    \caption{The visualization of decision polytopes represents the knowledge learned through training (a) the receiver\textquoteright s neural network, and (b) the sender\textquoteright s neural network. The directions of the arrows in the polytopes are the output decisions for a corresponding region of states, which are used to control the movement of the shepherd. The polytopes in green are the ones that are sent to the receiver.}
\label{fig:polytopes}
\end{figure*}

The framework extracts interpretable, high-fidelity decision rules from the neural networks with our EC-DT algorithm. These rules can be converted into a visual explanation of the knowledge learned by the original neural network, which is the mapping between the regions of input in the state space and the corresponding outputs. The visualization of receiver and sender\textquoteright s polytopes along with the output decisions of the shepherding agents are illustrated in Figure~\ref{fig:polytopes}. In previous literature~\cite{strombom2014solving}, the control force in shepherding task is a function of the direction of the sheep\textquoteright \ GCM relative to the shepherd and the direction of the target relative to the position of sheep\textquoteright \ GCM. Such a visualization provided in this paper might help domain experts to understand how the decisions of the models are made, and provide a way to inspect the knowledge learned models. Users can also see which pieces of knowledge from the sender have higher values and are fused to the receiver.

\begin{figure}
    \centering
    \includegraphics[width=0.9\linewidth]{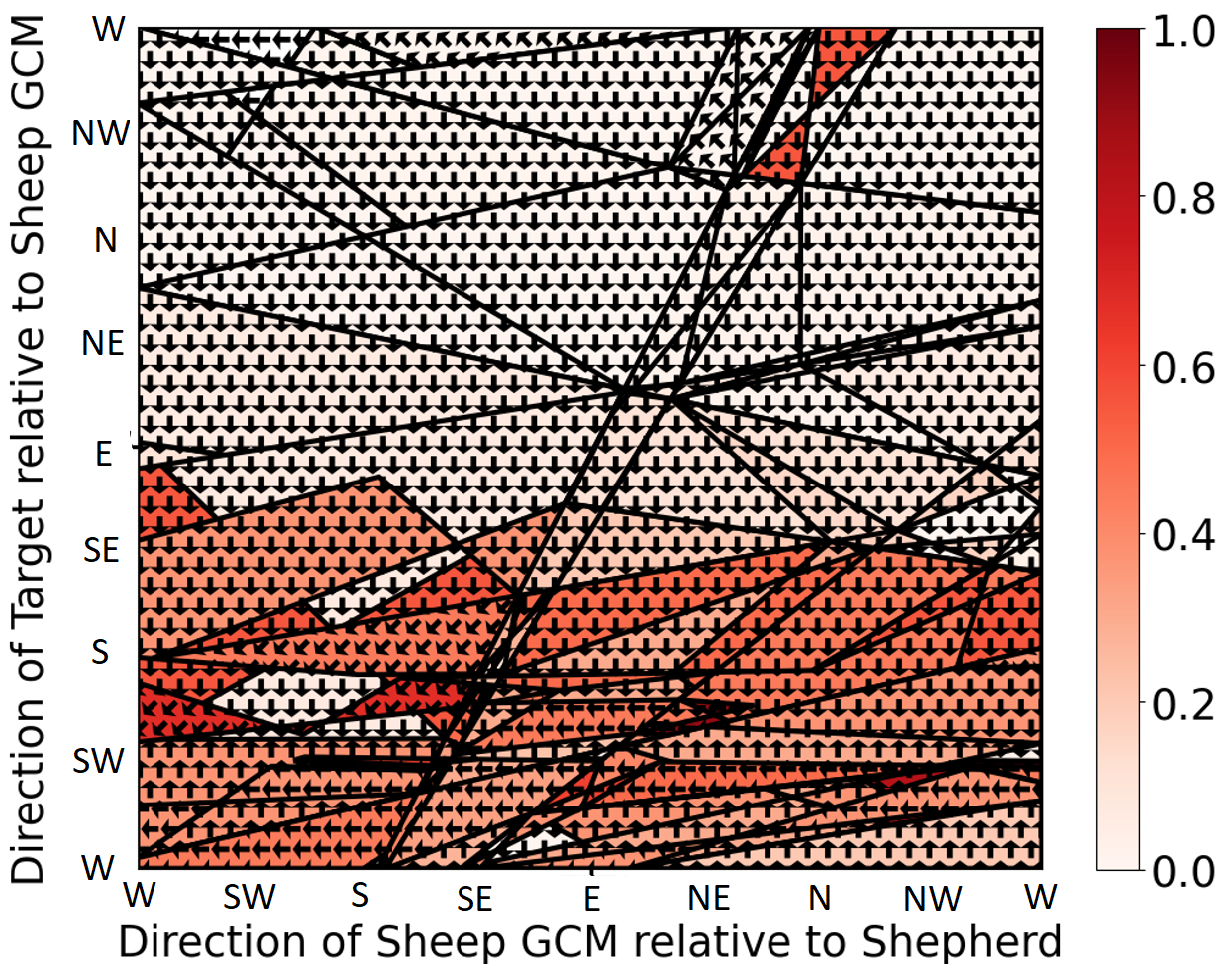}
    \caption{Visualization of weak state areas in the state space. The sheep\textquoteright \ GCM is located at 20 and 80 metres away from the shepherd and the target position. The stronger the red colour, the weaker the piece of knowledge represented by the corresponding polytope.}
    \label{fig:weak_state_area}
\end{figure}

Regarding the step involving the assessment of knowledge against the new environment, the level of the weakness of the pieces of knowledge can also be recorded and visualised. Figure~\ref{fig:weak_state_area} shows the fused rule set after knowledge fusion. The strength of the colour of each polytope indicates the degree to which the piece of corresponding knowledge is responsible for the failure of the task. For example, there are two polytopes with the strongest red colours near the lower-left corner of Figure~\ref{fig:weak_state_area}. These polytopes correspond to the case where the sheep are to the south or southwest of the shepherd, and the target is in the southwest direction of the sheep. In this case, there is an obstacle in the southwest direction according to the structure of environment C. The decision of the models according to the knowledge represented here is to guide the sheep in the southwest direction, which causes the sheep to collide with the obstacle. The surrounding polytopes have lighter colours which indicate that these pieces of knowledge are indirectly responsible for the failures by guiding the agents from low-risk states to high-risk ones. The visualization of the degree of weakness of the polytopes here also demonstrates the distribution of the priority factors for retraining and exploration used in the proposed framework.

\section{Conclusion}\label{jpaper2-conclusion}

In this paper, we propose an interpretable knowledge fusion framework for neural-based learning agents. The framework operates on sets of interpretable rules from the sender and receiver agents extracted from their original neural networks. The knowledge of both agents is evaluated in the new problem and compared to each other before the receiver decides to adopt which rules that would benefit it from learning the new task. The newly integrated rules are then converted into neural substrates and fused into the original neural network according to two proposed types of fusion: one with an increase in the size of a single hidden layer (type-1) and another with an increase in the depth of the model (type-2). The evaluation phase also determines the priority factors for sampling more instances or exploring the weaker region of the knowledge more frequently. The framework also allows the agent with a new knowledge structure to retrain to generalize on novel environments.

Performance on a synthetic binary classification problem demonstrates that the use of the type-1 fusion method for the new neural network provides more robust performance, higher interpretability, and non-redundant knowledge representation compared to the type-2 method. Type-1 neural network fusion is therefore more appropriate to use with our interpretable knowledge fusion framework between neural-based learning agents.

We proposed a retraining method called \textit{Prioritizing on Weak State Areas} (PoWSA), which increases the sampling rate of data in areas in the problem space where the agent does not perform well and assigns a higher exploration rate to these areas compared to other regions. This method accelerates retraining with new knowledge structures to generalize to novel environments. Even though the proposed method is $14.5\%$ more computationally expensive than the MULTIPOLAR method, it is more interpretable than MULTIPOLAR and all other baseline methods with no loss of task performance. The proposed framework also provides a way to generate a visual explanation of the knowledge from the neural network, represented by decision polytopes. The visualization may assist domain experts to understand and inspect the decisions made by the models, as well as determining which parts of the model or pieces of knowledge do not perform well in a task. The weak pieces of knowledge can also be visualized at different degrees so that the prioritized retraining process is more transparent to the users.

The primary limitation of the current framework is the lack of a mechanism to automatically analyse and identify which agents trained from previous environments are more appropriate to use for knowledge fusion, as there are no metrics employed to evaluate the complexity and resemblance of different environments an agent is exposed to. This is the focus of our future work in this area, where we will explore different complexity metrics, although we expect these to be task-specific. Moreover, while we tested the framework on a reinforcement learning task, the testing may expand to supervised learning tasks.

%


\bibliographystyle{IEEEtran}
\bibliography{references_NN-transfer}

\vspace{-1.5em}
\begin{IEEEbiography}[{\includegraphics[width=1in,height=1.25in,clip,keepaspectratio]{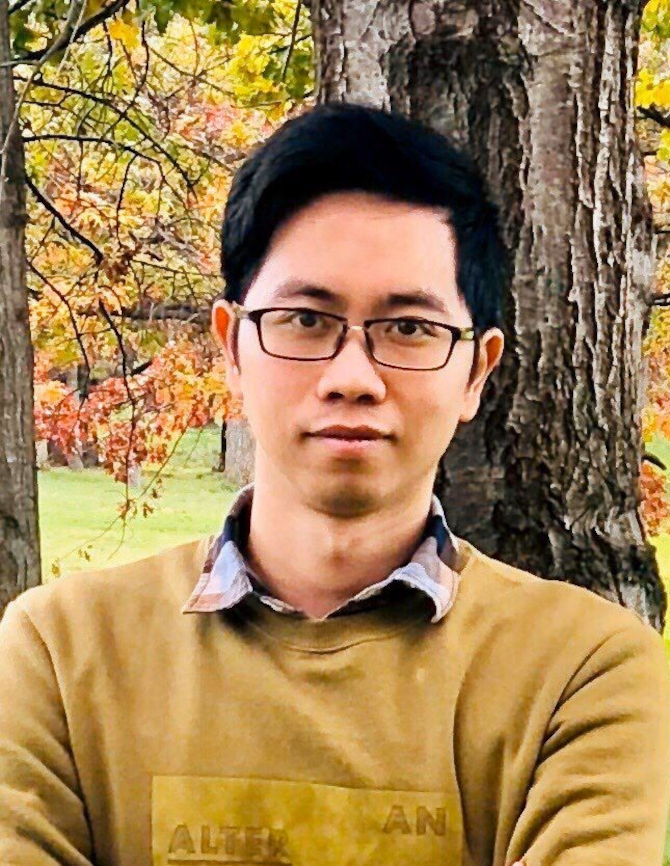}}]{Duy Tung Nguyen} 
(S\textquoteright 16) is pursuing a PhD at the University of New South Wales, Canberra, Australia. He received his B.Eng. degree in Biomedical Engineering from Hanoi University of Science and Technology, Vietnam in 2015 and his MSc in Computer Science from the University of New South Wales - Canberra in 2018. His research interests include deep learning, reinforcement learning, agent architecture, trusted autonomous systems, human-machine teaming, and interpretable and explainable artificial intelligence.
\end{IEEEbiography}

\vspace{-1.5em}

\begin{IEEEbiography}[{\includegraphics[width=0.9in,height=1.2in,clip,keepaspectratio]{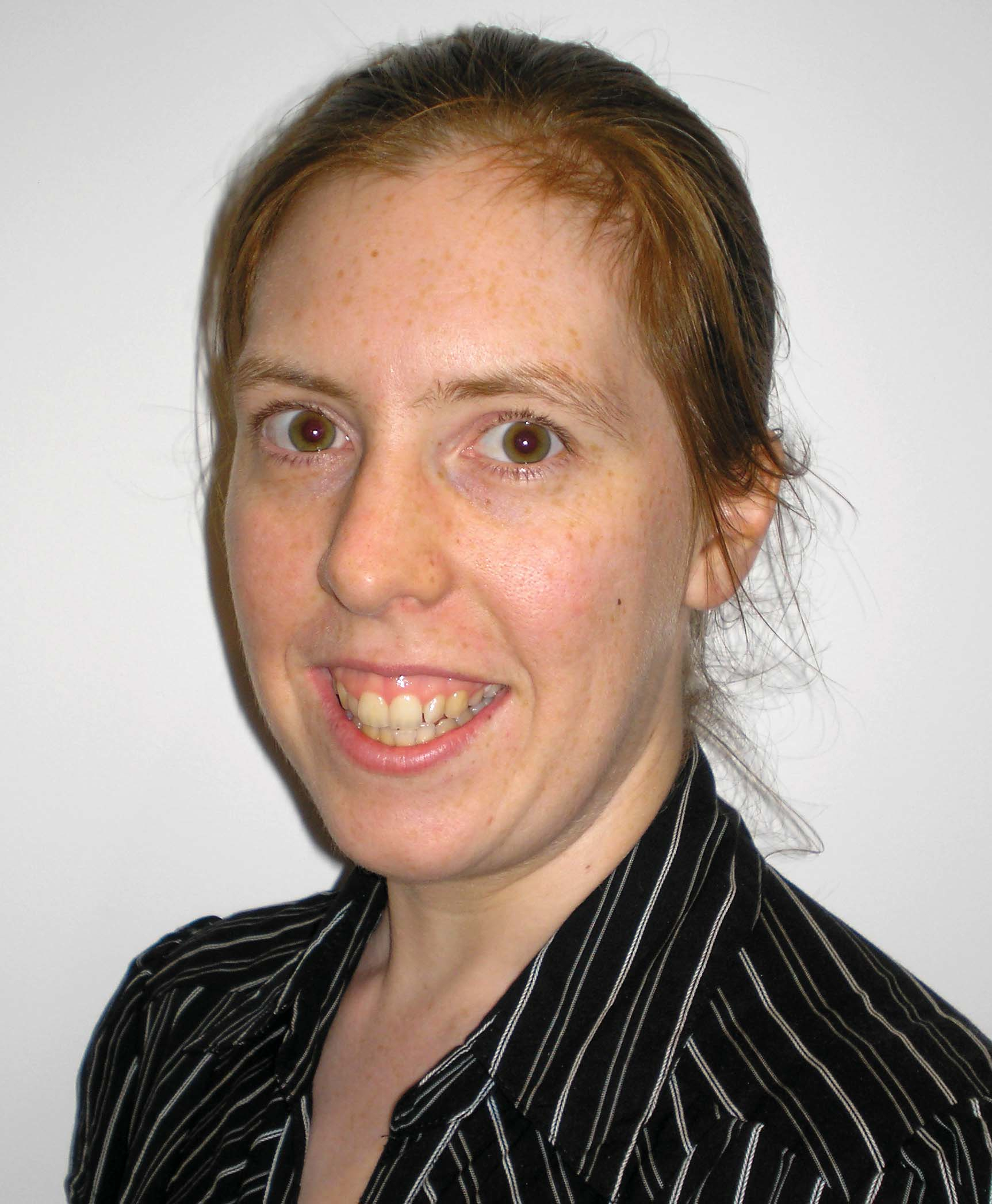}}]{Kathryn Kasmarik}
(SM\textquoteright 16) has a Bachelor of Computer Science and Technology (Advanced, Honours I, University Medal), University of Sydney, NSW, Australia, 2002; PhD (computer science), National ICT Australia and University of Sydney, NSW, Australia, 2007. She is a Professor in Computer Science at UNSW-Canberra, Australia. Her research lies in the areas of autonomous mental development and computational motivation, with applications in virtual characters, developmental robotics and intelligent environments.
\end{IEEEbiography}

\vspace{-1.5em}

\begin{IEEEbiography}[{\includegraphics[width=0.9in,height=1.2in,clip,keepaspectratio]{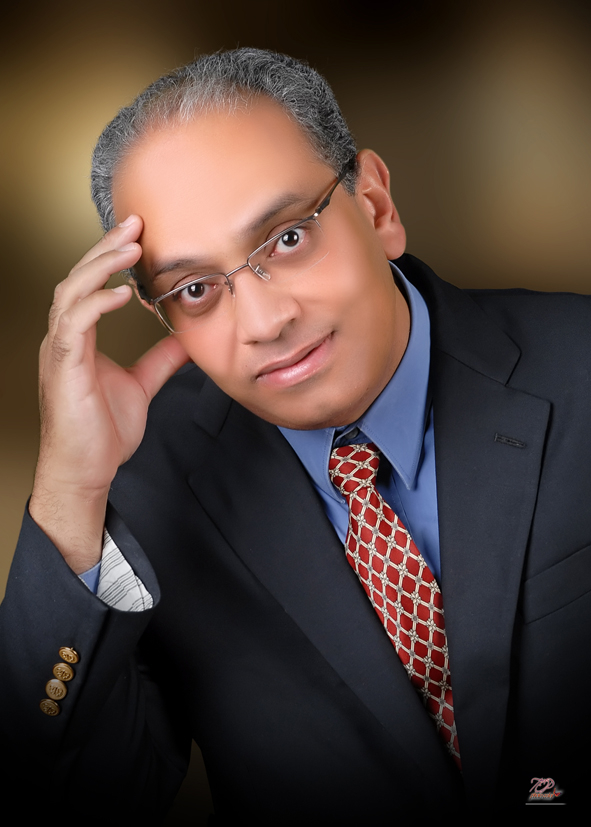}}]{Hussein Abbass}
(F\textquoteright 20) is a Professor at the University of New South Wales Canberra, Australia. He is the Founding Editor-in-Chief of the IEEE Transactions on Artificial Intelligence. He was the Vice-president of Technical
Activities (2016-2019) for the IEEE Computational Intelligence Society. His current research contributes to trusted human-swarm teaming with an aim to design next-generation trusted and distributed artificial intelligence systems that seamlessly integrate humans and machines.
\end{IEEEbiography}

\vfill

\end{document}